\documentclass[usenatbib,usegraphicx]{mn2e}
\usepackage{amsmath}
\usepackage{graphicx}
\usepackage{color}
\usepackage{units}
\usepackage{ulem}
\usepackage[usenames, dvipsnames, svgnames, table]{xcolor}
\DeclareGraphicsExtensions{.eps,.pdf,.png,.jpg}

\makeatletter
 \def\@textbottom{\vskip \z@ \@plus 1pt}
 \let\@texttop\relax
\makeatother

\begin{document}

\title[Disc Evolution]{Towards a radially-resolved semi-analytic model for the evolution of disc galaxies tuned with machine learning}

\author[J. C. Forbes et al.]{John C. Forbes,$^1$\textsuperscript{\thanks{E-mail: john.forbes@cfa.harvard.edu}}  Mark R. Krumholz,$^{2,3}$ Joshua S. Speagle$^{1,4}$\\ 
$^1$ Harvard-Smithsonian Center for Astrophysics, 60 Garden Street, Cambridge, MA USA\\
$^2$ Research School of Astronomy \& Astrophysics, Australian National University, Canberra, ACT 2611, Australia\\
$^3$ Centre of Excellence for Astronomy in Three Dimensions (ASTRO-3D), Australia \\
$^4$ NSF Graduate Research Fellow
}

\maketitle

\begin{abstract}
We present a flexible, detailed model for the evolution of galactic discs in a cosmological context since $z\approx 4$, including a physically-motivated model for radial transport of gas and stars within galactic discs. This expansion beyond traditional semi-analytic models that do not include radial structure, or include only a prescribed radial structure, enables us to study the internal structure of disc galaxies and the processes that drive it. In order to efficiently explore the large parameter space allowed by this model, we construct a neural network-based emulator that can quickly return a reasonable approximation for many observables we can extract from the model, e.g. the star formation rate or the half mass stellar radius, at different redshifts. We employ the emulator to constrain the model parameters with Bayesian inference by comparing its predictions to 11 observed galaxy scaling relations at a variety of redshifts. The constrained models agree well with observations, both those used to fit the data and those not included in the fitting procedure. These models will be useful theoretical tools for understanding the increasingly detailed observational datasets from IFUs.
\end{abstract}


\section{Introduction}

The basic story of how galaxies form and evolve is quite similar to the vision laid out in classic papers nearly four decades ago \citep{ white_core_1978, fall_formation_1980, blumenthal_formation_1984}, wherein gas cools in the potential wells of cold dark matter halos. Galaxy formation modelling has also had extensive success deriving a detailed array of inferences about galaxies using the complementary tools of full hydrodynamic simulations \citep[e.g.][]{hopkins_galaxies_2013, vogelsberger_introducing_2014, pillepich_simulating_2018, schaye_eagle_2015, ceverino_high-redshift_2010} and dark matter-only simulations plus empirical modeling \citep[e.g.][]{behroozi_average_2013, hearin_dark_2013}.  
Despite all of the success, the fundamental physics controlling the evolution of galaxies remains poorly understood, from galactic winds to star formation to quenching. Given these unknowns, it is clear that many aspects of galaxy formation need to be parameterized. Models that do so explicitly and with sufficient speed to economically survey the space of plausible parameter values provide the potential to narrow constraints on the parameters in question and guide future observational efforts.

There is a long history of such models constructed independently with a variety of strengths and weaknesses \citep[e.g.][]{benson_what_2003, bower_breaking_2006, cattaneo_modelling_2006, somerville_semi-analytic_2008, guo_dwarf_2011, lu_semi-analytic_2014, henriques_galaxy_2015}. Typically these models are constructed explicitly on a foundation of halo merger trees and treat the evolution of the galaxies themselves in a reasonably simple way. The parameters are then adjusted to fit the luminosity function of galaxies in various bands at various redshifts to the degree possible. Quantities that depend on the radial distribution of material in the galaxy can often only be inferred by way of strong implicit or explicit assumptions that are unlikely to be true in the general case.

Thanks to extensive observational efforts in the past decade, a rich set of data are available that extend beyond luminosity functions. Large populations of galaxies have been shown to follow numerous strong correlations between stellar mass and a wide array of other properties: gas fractions, star formation rates, metallicity of gas and stars, physical size, central stellar surface densites, Sersic index, angular momenta, and circular velocity. In parallel, new relationships have been uncovered at the kpc scale by IFU surveys of nearby galaxies \citep{sanchez_califa_2012, croom_sydney-aao_2012, bundy_overview_2015}

In order to compare theories to data at scales smaller than a galaxy, including IFU data, and profiles of galaxies at all redshifts, e.g. measurements of galaxy size, concentration, and more generally star formation and density profiles \citep[e.g.][]{lang_bulge_2014, tacchella_evolution_2016, bigiel_universal_2012, nelson_where_2016}, it is necessary to include some spatial dependence in one's model of galaxy evolution. Obviously cosmological hydrodynamic simulations meet this requirement, but are generally too expensive to efficiently explore the wide range of plausible parameter values. One-dimensional simulations strike a good balance between the variety of quantities they can model and the speed necessary to do so efficiently. A variety of models have been developed along these lines aiming to fit one or another particular observational relation \citep[e.g.][]{van_den_bosch_impact_2002, dutton_impact_2009, fu_star_2013, stevens_building_2016}.

In this paper we present a new one-dimensional model based on the Gravitational Instability Dominated Galaxy Evolution Tool \citep[GIDGET;][]{forbes_evolving_2012, forbes_balance_2014}. Our model includes a comprehensive set of parameterized physical processes, some of which are similar to those included in many semi-analytic models (SAMs), including cosmological accretion and halo growth, metallicity-dependent molecular gas fractions and star formation, galactic outflows that depend on star formation rate and other galactic properties, the production, expulsion via winds, and delayed return of $\alpha$ and iron peak elements, and realistically-delayed gas return from older stellar populations. However, our model also includes a number of ingredients that have not heretofore been included in SAMs, but which both theory and observation suggest are important to the structure of real galaxies: the radial transport of gas and stars via gravitational and magnetic torques, and the role of this transport in regulating the galaxy rotation curve, turbulence in the interstellar medium, and the radial distribution of metals. We aim to constrain the parameters describing all of these processes, and to quantify the importance of each of them in setting observable quantities, using a novel machine learning framework.
 
In this paper we present a reasonably sophisticated one-dimensional model. We include a comprehensive set of parameterized physical processes, and we aim to constrain these parameters, and quantify the importance of each in setting individual observable quantities. In particular we include prescriptions for cosmological accretion and halo growth, metallicity-dependent molecular gas fractions and star formation, extremely flexible spatially-dependent galactic outflows, the production, diffusion, advection, and fountain-related mixing of $\alpha$ and iron peak elements, realistically-delayed return of gas from intermediate-age and old stellar populations, radial transport of gas and stars via viscous torques, the driving of turbulence via these torques and stellar feedback, and an evolving rotation curve calculated from the instantaneous mass distribution.

Section \ref{sec:model} explains these prescriptions in detail, section \ref{sec:search} details the procedure used to compare the model to a wide array of observational relations, and section \ref{sec:results} shows the fits thereby obtained, and a quantification of the importances of each model parameter in setting each observational quantity. We summarize in section \ref{sec:summary}.

\section{The Model}
\label{sec:model}

To model a single galaxy as a function of time, we employ the latest version of GIDGET, \citep{forbes_evolving_2012, forbes_balance_2014}. At its most basic level, GIDGET solves the full equations of hydrodynamics and stellar dynamics for the evolution of a thin, axisymmetric viscously-evolving disc. In the code, a galactic disc is discretized onto a radial grid, with the gas and stellar column densities, velocity dispersions, and metallicities tracked in every annulus. These quantities are evolved forward in time under the assumptions that the disc is axisymmetric ($\partial/\partial\phi=0$, i.e. no quantities are allowed to vary with azimuthal coordinate $\phi$), and thin ($|v_r|\ll \sigma \ll v_\phi$, i.e. the magnitude of the velocity of any bulk radial motions $|v_r|$ must be smaller than the velocity dispersion $\sigma$, which itself must be small compared to the azimuthal velocity $v_\phi$), so that each of the quantities in question is a function of radius and time only. 

In the remainder of this section we describe the GIDGET model for galaxy evolution. We give the evolution equations for the gas and stars in a galaxy in section 2.1. These equations depend on a series of terms, which we explain in the subsequent sections. In section 2.2 we describe how we compute the galactic rotation curve and its evolution. In 2.3 we present our model for gravitational instability-driven transport of gas and stars within galaxies. Section 2.4 explains our model for cosmological accretion, and sections 2.5 and 2.6 describe how we implement star formation and stellar feedback. Throughout, we provide a full derivation or explanation of the method only in places where our method here differs from that in \citet{forbes_balance_2014}. For other parts of the method, we refer readers to that paper and to \citet{forbes_evolving_2012}. Physical and numerical parameters introduced throughout this section are summarized in Table \ref{tab:priors}.

\subsection{Hydrodynamics}
\label{sec:hydro}
The surface density of the gas ($\Sigma$) and the stars ($\Sigma_*$) are tracked following the standard continuity equation in $n_x$ logarithmically spaced annuli between $r_\mathrm{min}$ and $r_\mathrm{max}$, so that excluding the source and sink terms, mass would be exactly conserved. 
\begin{eqnarray}
\label{eq:dcoldt}
\frac{\partial\Sigma}{\partial t} &=& \frac{1}{2\pi r} \frac{\partial\dot{M}}{\partial r} + \dot{\Sigma}_\mathrm{cos} - (f_{R,\mathrm{inst}}+\mu) \dot{\Sigma}_\mathrm{SF} + \dot{\Sigma}_\mathrm{rec} \\
\frac{\partial\Sigma_*}{\partial t} &=& \frac{1}{2\pi r} \frac{\partial\dot{M}_*}{\partial r} + f_{R,\mathrm{inst}} \dot{\Sigma}_\mathrm{SF} - \dot{\Sigma}_\mathrm{rec}
\end{eqnarray}
Mass moves between the annuli at a rate $\dot{M}$ for the gas and $\dot{M}_*$ for the stars, with positive values indicating inward flow. These quantities are non-trivial functions of radius and time, and will be discussed in more detail in section \ref{sec:transport}. Mass is added to the gaseous component at each annulus via $\dot{\Sigma}_\mathrm{cos}$, representing cosmological accretion, which again depends on radius and time (see section \ref{sec:acc}). The star formation rate surface density at each radius is given by $\dot{\Sigma}_\mathrm{SF}$. Of the surface density of gas which forms stars each time step, only some fraction $f_{R,\mathrm{inst}} \approx 0.77$ remains in long-lived stellar remnants -- the remainder is returned to the ISM via core collapse supernovae on timescales short enough that we approximate them as instantaneous following \citet{tinsley_evolution_1980}. Low and intermediate mass stars also return mass to the ISM via stellar winds on much longer timescales. This is included via the term $\dot{\Sigma}_\mathrm{rec}$, which we take to be the following summation over each bin of stellar age tracked by the simulation each with age $T_i$,
\begin{equation}
\dot{\Sigma}_\mathrm{rec} = \sum_i \Sigma_{*,i} \cdot \begin{cases}
 0 \ \ \ \ \ \ \ \ \ \ \ \ \ \ \ \ \ \mathrm{if}\ T_i<40\ \mathrm{Myr} \\
 d f_{ml}(T_i)/d T_i \ \ \mathrm{otherwise} \\
 \end{cases}
\end{equation}
where, following \citet{leitner_fuel_2011}, the fraction of mass returned to the ISM from a mono-age stellar population of age $T$ is taken to be
\begin{equation}
f_{ml}(T) = 0.046\cdot \ln\left( \frac{T}{2.76 \times 10^5\ \mathrm{yr}} + 1 \right)
\label{eq:fml}
\end{equation}
assuming a \citet{chabrier_galactic_2003} IMF and the functional form of \citet{jungwiert_continuous_2001}. Finally, mass is ejected from the disc permanently at a rate proportional to the local star formation rate, where the ratio of the outflow rate to the star formation rate is defined as the mass loading factor $\mu$ (see section \ref{sec:metals}).

The gas velocity dispersion, including both internal and turbulent kinetic energy is evolved according to,
\begin{equation}
\frac{\partial \sigma}{\partial t} = \frac{\sigma}{6\pi r\Sigma}\frac{\partial}{\partial r}\dot{M} + \frac{5(\partial \sigma/\partial r)}{6\pi r \Sigma} \dot{M} + \frac{(\beta-1) v_\phi}{6\pi r^3 \Sigma \sigma}\mathcal{T} + \frac{\mathcal{G}-\mathcal{L}}{3\sigma\Sigma} 
\end{equation}
We use $\beta= d\ln v_\phi/d\ln r$ to denote the local power law index of the rotation curve. The first two terms account for advection of kinetic plus internal energy through the disc. The third term represents viscous heating via local torques $\mathcal{T}$, and the final term accounts for local heating and cooling, with the net energy gain rate per unit surface area $\mathcal{G}-\mathcal{L}$. This rate is taken to be
\begin{equation}
\label{eq:netheating}
\mathcal{G} - \mathcal{L} = \chi_\mathrm{inj}\left\langle\frac{p}{m_*}\right\rangle \sigma \dot{\Sigma}_\mathrm{SF} -\eta \Sigma \sigma^2 \kappa Q_g^{-1} \left( 1 + \frac{\sigma \Sigma_*}{\sigma_\mathrm{zz}\Sigma}\right) \left(1 - \frac{\sigma_\mathrm{sf}^2}{\sigma^2}\right)^{3/2}
\end{equation}
The first term accounts for energy added to the gas by supernova remnants as the momentum they acquired during the Sedov phase is deposited into the ISM. The mean momentum added per unit stellar mass formed is taken to be the standard value of 3000 km/s, and $\chi_\mathrm{inj}$ is a free parameter. The second term, excluding the final factor, is simply the kinetic energy per unit area $(3/2) \Sigma \sigma^2$ divided by a scale height crossing time, replacing the factor of 3/2 with a free parameter $\eta$. This is based on the classic result that turbulent kinetic energy decays in a crossing time \citep{stone_dissipation_1998, mac_low_kinetic_1998}. The final factor truncates the cooling as $\sigma \rightarrow \sigma_\mathrm{sf}$. Once the velocity dispersion approaches the gas temperature of the WNM, the large-scale turbulence no longer dominates the energy, and the velocity dispersion is likely set by a balance of heating and cooling \citep[e.g.][]{wolfire_neutral_2003}, which we take to be a free parameter $\sigma_\mathrm{sf}$.

Stars in the disc are also subject to transport, and hence a process analogous to viscous heating. Moreover, as new stars are added to the disc, they form with a velocity dispersion comparable to that of the gas from which they form, which tends to decrease the velocity dispersion of the overall population. We separately track the radial ($\sigma_{\rm rr}$) and vertical ($\sigma_{\rm zz}$) velocity dispersions of the stars. The former evolves according to
\begin{eqnarray}
\frac{\partial \sigma_\mathrm{rr}}{\partial t} = \frac{1}{2\pi r\Sigma_* (\sigma_{rr}+\sigma_{zz})}\bigg\{ \frac{v_\phi (\beta-1)}{r^2} \mathcal{T}_* + \sigma_{rr}^2 \frac{\partial \dot{M}_*}{\partial r} + \\
 \dot{M}_* \left(3\sigma_{rr}\frac{\partial \sigma_{rr}}{\partial r} + 2\sigma_{zz}\frac{\partial \sigma_{zz}}{\partial r}\right)\bigg\}  + \frac{1}{2 \Sigma_* \sigma_{rr}} \dot{\Sigma}_\mathrm{SF} \sigma^2   \nonumber
\end{eqnarray}
The set of terms in braces encapsulates the viscous heating and transport as derived from the Jeans equations in \citet{forbes_balance_2014}, while the final term accounts for the addition of new stars to the population with the velocity dispersion of the gas, $\sigma$. As in \citet{forbes_balance_2014}, we assume that viscous heating has a lesser effect on the vertical velocity dispersion of the stars $\sigma_{zz}$ than the in-plane velocity dispersion $\sigma_{rr}$ by a factor of two, so that 
\begin{eqnarray}
\frac{\partial \sigma_\mathrm{zz}}{\partial t} =\frac12 \frac{1}{2\pi r\Sigma_* (\sigma_{rr}+\sigma_{zz})}\bigg\{ \frac{v_\phi (\beta-1)}{r^2} \mathcal{T}_* + \sigma_{rr}^2 \frac{\partial \dot{M}_*}{\partial r} + \\
 \dot{M}_* \left(3\sigma_{rr}\frac{\partial \sigma_{rr}}{\partial r} + 2\sigma_{zz}\frac{\partial \sigma_{zz}}{\partial r}\right)\bigg\}  + \frac{1}{2 \Sigma_* \sigma_{zz}} \dot{\Sigma}_\mathrm{SF} \sigma^2   \nonumber
\end{eqnarray}
For numerical stability, we also never allow the velocity dispersion of the stars to drop below $\sigma_{*,\mathrm{min}} = 10\ \mathrm{km}\ \mathrm{s}^{-1}$.

\subsection{Rotation Curve}
\label{sec:rot}
In contrast to previous work with GIDGET, we do not assume that the circular velocity is constant in time. Instead, we self-consistently calculate it based on the distribution of matter ($\Sigma(r)$ and $\Sigma_*(r)$) in the disc, and a model for $\rho_\mathrm{DM}$ as a function of redshift, halo mass, and deviation from the redshift-dependent concentration-halo mass relation.

The circular velocity\footnote{Note that in the derivation of the dynamical equations of the previous section, we have neglected asymmetric drift, i.e. the difference between the circular velocity set by the gravitational potential and the mean tangential component of the velocity $v_\phi$.} at every point in the disc can be divided into contributions from the bulge, the dark matter halo, and the self-gravity of the disc respectively,
\begin{equation}
v_\phi^2 = v_{\phi,b}^2 + v_{\phi,\mathrm{dm}}^2 + v_{\phi,\mathrm{disc}}^2.
\end{equation} 
In this context, the bulge is the material inside the inner cutoff of our logarithmic grid. The bulge mass grows over time under the assumption that any gas which arrives there by in-disc transport or directly via cosmological accretion rapidly forms stars, i.e.
\begin{equation}
\label{eq:mdotcentral}
\dot{M}_\mathrm{central} = \dot{M}_*\big|_{r_0} + (\dot{M}\big|_{r_0}+\dot{M}_\mathrm{acc}(r<r_0)) \frac{f_{R,\mathrm{asym}}}{f_{R, \mathrm{asym}}+\mu}
\end{equation}
Here $\mu$ is the mass loading factor evaluated at the innermost cell of the simulation, and $\dot{M}$ and $\dot{M}_*$ are the in-disc transport rates of gas and stars respectively, as calculated in the following section. The accretion rate within the inner radius of the domain, $r_0$, is $\dot{M}_\mathrm{acc}(r<r_0) = \int_0^{r_0} \dot{\Sigma}_\mathrm{acc} 2\pi r dr $. Mass added to the bulge via gas, both in-disc and from cosmological accretion, is reduced by a factor $f_{R,\mathrm{asym}}/(f_{R,\mathrm{asym}}+\mu)$, which is the long-run fraction of some initial gas mass that survives in long-lived stellar remnants. The asymptotic remnant fraction, $f_{R,\mathrm{asym}}$ is given by $1-f_{ml}(13.7\ \mathrm{Gyr}) \approx 0.503$ according to Equation \eqref{eq:fml}. The contribution to the circular velocity from this central ``bulge'' material is easily computed as $v_{\phi,b}^2 = G M_\mathrm{central} / r$.

The contribution from the dark matter halo is similarly straightforward, namely $v_{\phi,\mathrm{dm}}^2 = G M_\mathrm{dm}(<r)/r$. To compute the dark matter mass interior to radius r, $M_\mathrm{dm}(<r)$, we need an explicit model of the halo density profile, including its dependence on redshift and halo mass. For an Einasto profile the mass interior to a given halo-centric radius is
\begin{equation}
M_\mathrm{dm}(<r) = \pi \rho_s r_s^3 2^{2-3/\alpha} e^{2/\alpha}  \alpha^{3/\alpha-1}   \Gamma\left(\frac3\alpha, \frac2\alpha\left(\frac{r}{r_s}\right)^\alpha\right)
\end{equation}
where $\Gamma(x,z)$ is the incomplete Gamma function $\int_0^z e^{-t} t^{x-1} dt$. The parameters $\alpha$, $r_s$, and $\rho_s$ must be set to fit the halos formed in a cosmological simulation. We adopt the results from the simulations of \citet{dutton_cold_2014} for the relationship between the halo mass and redshift and the other parameters, allowing some offset in the mass-concentration relation $\alpha_\mathrm{con}$,
\begin{eqnarray}
\label{eq:alphacon}
c_{200} &=& 10^{a+b m + \alpha_\mathrm{con}}\\
\alpha &=& 0.0095 \nu^2 + 0.155
\end{eqnarray} 
where the new constants $a$, $b$, $m$, and $\nu$ are given by the following
\begin{eqnarray}
m &=& \log_{10} \left( \frac{M_h h}{10^{12} M_\odot} \right)\\
\nu &=& 10^{-0.11 + 0.146 m + 0.0137 m^2 + 0.00123 m^3} \\
& & \ \ \ \ \times \ \ \  \left(0.033 + 0.79(1+z) + 0.176 e^{-1.356z}\right) \nonumber \\
a &=& 0.520 + (0.905-0.520) \exp{\left(-0.617z^{1.21}\right)}\\
b &=& -0.101 + 0.026 z
\end{eqnarray}
The value of $\rho_s$ is set such that the dark matter mass within $R_\mathrm{Vir}$ is in fact $M_h$, 
\begin{equation}
\rho_s =  M_h \left[ \pi \rho_s r_s^3 2^{2-3/\alpha} e^{2/\alpha}  \alpha^{3/\alpha-1}   \Gamma\left(\frac3\alpha, \frac2\alpha c_{200}^\alpha\right) \right]^{-1}
\end{equation}

Finally, for the contribution to the rotation curve from the self-gravity of the disc, we follow \citet{binney_galactic_2008},
\begin{equation}
v_{\phi,\mathrm{disc}}^2 = -r \int_0^\infty S(k) J_1(k r) k dk,
\end{equation}
where $J_i(x)$ denotes the $i$th Bessel Function of the first kind, and 
\begin{equation}
S(k) = -2\pi G \int_0^\infty J_0(k r) (\Sigma(r)+\Sigma_*(r)) r dr.
\end{equation}
For the purposes of our simulations, we approximate the disc surface density as piecewise-constant in each annulus. In so doing, we can re-write these equations as
\begin{eqnarray}
\label{eq:vphidisc}
v_{\phi, \mathrm{disc}}^2 &=& 2 \pi r G \times \\
\nonumber & & \sum_i (\Sigma_i+\Sigma_{*,i}) \int_0^\infty dk  J_1(k r) k \int_{r_{i-1/2}}^{r_{i+1/2}} J_0(k r') r' dr'
\end{eqnarray}  
where $\Sigma_i$ and $\Sigma_{*,i}$ are the gas and stellar surface densities in the $i$th annulus of the simulation, and $r_{i\pm 1/2}$ are the boundaries of the annulus. Details of how these integrals are computed may be found in Appendix \ref{app:rotcurve}.

\subsection{Gravitational Instability}
\label{sec:transport}
When cells in the disc have values of the multi-component \citet{toomre_gravitational_1964} $Q$ value (defined in \citet{romeo_effective_2011}) that fall below $Q_f$, a free parameter, they experience a torque which will tend to heat and transport the gas. The physical motivation is that when $Q$ is low enough the disc will be subject to local gravitational instability, which will induce turbulent heating whose ultimate source is the gravitational potential of the galaxy \citep{bournaud_ism_2010, krumholz_dynamics_2010, goldbaum_mass_2015, goldbaum_mass_2016, behrendt_clusters_2016}. 

To encapsulate this effect, we have used a set of basic equations that includes viscous heating terms proportional to, $\mathcal{T}$, $\partial\mathcal{T}/\partial r \propto \dot{M}$, and $\partial^2\mathcal{T}/\partial r^2 \propto \partial \dot{M}/\partial r$. The torque and the mass flux are related via
\begin{equation}
\dot{M} = \frac{-1}{v_\phi (1+\beta)} \frac{\partial \mathcal{T}}{\partial r},
\end{equation}
which is a statement of the conservation of angular momentum under the assumption of a slowly-varying potential \citep{krumholz_dynamics_2010}. The torque $\mathcal{T}$ is set to zero in annuli where $Q>Q_f$. When $Q<Q_f$, we compute the torque (and its derivatives) as a solution to a simple boundary value problem such that 
\begin{equation}
\label{eq:dQdt}
dQ/dt = {(Q_f-Q)v_\phi/r} \ \ \mathrm{if}\ Q<Q_f.
\end{equation} In other words, we assume that turbulence driven by gravitational instability acts to return the disc to a marginally stable state on a dynamical timescale, and that more unstable discs experience quicker reversion to $Q=Q_f$. In addition to whatever torque is obtained by solving this boundary value problem, a torque given by
\begin{equation}
\label{eq:alphaMRI}
\mathcal{T}_\mathrm{MRI} = 2\pi \alpha_\mathrm{MRI} \Sigma r^2 \sigma_\mathrm{sf}^2 
\end{equation}
is added following the $\alpha$ prescription of \citet{shakura_black_1973}. This accounts for the possibility that radial transport in the disc may occur through some other mechanism besides gravitational instability, though for simplicity we assume $\alpha_\mathrm{MRI}$ is constant in time and space within a given model.

The stars experience an analogous set of torques and radial derivatives thereof, denoted $\mathcal{T}_*$, $\dot{M}_* \propto \partial\mathcal{T}_*/\partial r$ and so forth. The stars conserve their specific angular momentum so that
\begin{equation}
\dot{M}_* = \frac{-1}{v_\phi (1+\beta)} \frac{\partial \mathcal{T}_*}{\partial r},
\end{equation}
as in \citet{forbes_balance_2014}. $\mathcal{T}_*$ is similarly set to zero when $Q_*>Q_\mathrm{lim}$, and to some non-zero value when $Q_*<Q_\mathrm{lim}$, such that $dQ_*/dt = (Q_\mathrm{lim} - Q_*) v_\phi/(4rQ_*)$, as suggested by \citet{sellwood_spiral_1984} and \citet{carlberg_dynamical_1985}. 

The heating mechanisms for the gas and stars are subtly different. Stars are heated according to their stability parameter $Q_* = \sigma_\mathrm{rr} \kappa/(\pi G \Sigma_*)$ where the epicyclic frequency $\kappa = \sqrt{2(\beta+1)}$, and $\beta=d\ln v_\phi/d\ln r$. In contrast, the torque experienced by the gas depends on 
\begin{equation}
Q \approx Q_\mathrm{RW} =
\begin{cases}
 \left(\frac{W}{Q_*} + \frac1{Q_g} \right)^{-1} & \mathrm{if} Q_*T_*>Q_g T_g \\
 \left(\frac{1}{Q_*} + \frac{W}{Q_g} \right)^{-1} & \mathrm{if} Q_*T_*<Q_g T_g\\
\end{cases}
\end{equation}
where the weight $W=2\sigma_\mathrm{rr}\sigma/(\sigma_\mathrm{rr}+\sigma)$,  and $T_*\approx 0.8+0.7 \sigma_\mathrm{zz}/\sigma_\mathrm{rr}$, and $T_g \approx 1.5$ are corrections for the finite thickness of each component. In other words the global stability of the disc, and hence the torque on the gas, depends on both the gas and stars, whereas the torques on the stars are assumed to only be affected by the stability of the stars, determined by $Q_*$ alone. In practice this means that we must choose $Q_\mathrm{lim} \ga Q_f$, since otherwise the disc may reach a state in which no torque on the gas $\mathcal{T}$ will be sufficient to return the disc to $Q=Q_f$. Intriguingly \citet{romeo_what_2016} have demonstrated that the core of a nearby galaxy is in just such a state, where the stars dominate the global gravitational instability of the disc, potentially driving substantial turbulence in the gas phase. This suggests that further work on the interplay between gravitational instability of different disc components, particularly their ability to drive torques on the other components, warrants further study.

\subsection{Accretion}
\label{sec:acc}
\subsubsection{Mean Accretion Rate}
Following \citet{bouche_impact_2010} the mean dark matter accretion rate for a halo of mass $M_h$ at a redshift $z$ is taken to be
\begin{equation}
\label{eq:dMhdt}
\frac{d M_h}{dt}\bigg|_\mathrm{avg}  = 34 (M_h/10^{12} M_\odot)^{1.14} (1+z)^{2.4}\ \ M_\odot/\mathrm{yr}.
\end{equation}
The numerical values of these exponents are somewhat uncertain, and there are alternative formulations for the halo growth rate \citep{van_den_bosch_universal_2002, wechsler_concentrations_2002, neistein_constructing_2008, genel_mergers_2008, mcbride_mass_2009}, but this formula captures the basic features of the growth of dark matter halos.

Baryons are accreted according to $\dot{M}_b = \epsilon_\mathrm{acc} f_b dM_h/dt$. Essentially, baryons are assumed to closely follow the dark matter. This assumption is useful for its simplicity, though there are reasons to expect it to fail. For instance, since $z=1$ most of the growth in halo mass has been due to the decrease in the mean density of the universe rather than new dark matter being added to halos \citep{diemer_pseudo-evolution_2013}. However, \citet{wetzel_physical_2015} have shown in cosmological hydrodynamic simulations that the growth of baryonic mass at small radii in a halo is actually closely coupled to the growth of halo mass, despite pseudo-evolution.

A fit for $\epsilon_\mathrm{acc}$ is provided by \citet{faucher-giguere_baryonic_2011} for $z>2$, but we continue to use it at lower redshift for simplicity,
\begin{equation}
\epsilon_\mathrm{acc} =\epsilon_\mathrm{reduce} \min\left(\epsilon_\mathrm{ceil},\ 0.47  \left(\frac{M_h}{10^{12} M_\odot}\right)^{-0.25}\left(\frac{1+z}{4}\right)^{0.38}\right)
\label{eq:acceff}
\end{equation}
We impose a maximum accretion efficiency $\epsilon_\mathrm{max} \le1$, which is a free parameter that acts to allow inefficient accretion even in low-mass halos. Including this parameter was required to fit $z=0$ galaxy scaling relations in the simple equilibrium models in \citet{forbes_origin_2014}, and may be plausibly explained as pre-heating in low-mass halos by \citet{lu_semi-analytic_2014, lu_formation_2015}, or mass lost from smaller galaxies which reside in the substructures accreted along with the smooth flow of dark matter. Equation \ref{eq:acceff} includes a factor $\epsilon_\mathrm{reduce}$, which will be specified in section \ref{sec:quenching} to account for quenching. This formula also captures a few important physical features of gas accretion onto halos, namely that higher-mass halos have higher Virial temperatures, so shock-heated gas cools more slowly. Similarly, higher-redshift halos are denser, so the cooling time is shorter, and the efficiency is higher. 

\subsubsection{Variability in the Accretion Rate}
Equation \ref{eq:dMhdt} is useful as a starting point for understanding the growth of halos, but obviously it does not capture halo-to-halo variance in the accretion history, nor the effects of mergers. To incorporate these elements into our model, we employ publicly-available halo merger trees from the Bolshoi simulation \citep{klypin_dark_2011, behroozi_gravitationally_2013}. For each tree, we construct the growth history of the main progenitor, and identify halos which merge into the main halo between each output snapshot of Bolshoi. The finite resolution of the N-body simulation means that halos below a certain mass are not formed in the simulation, and that not all $z=0$ halos have a main progenitor at the higher redshifts where we begin our 1D calculations, e.g. $z=4$.

In order to get around this resolution limit, we rely on the fact that the growth of structure in a cold dark matter universe is self-similar below a characteristic halo mass $\mathcal{M}^*$. Rather than following the growth of a galaxy in a particular halo with a particular final mass in Bolshoi, we scale each main progenitor growth history so that it can be used to simulate a galaxy in a halo of any mass, though in practice we restrict ourselves to use trees whose final halo mass is within 0.25 dex of the halo we wish to simulate if possible. If we wish to simulate a halo whose mass is below the resolution limit, we pick a tree from the lowest half dex of available trees. 

To do so, we simply take the instantaneous mass and redshift of the Bolshoi main progenitor halo, use equation \ref{eq:dMhdt} to find the average dark matter accretion rate for halos at that mass and redshift, and record $x$ such that
\begin{equation}
\label{eq:randomfactors}
10^x =  \min\left( \frac{dM_h/dt |_\mathrm{Bolshoi}}{dM_h/dt(M_{h,\mathrm{Bolshoi}},z) |_{\mathrm{avg} }}, 10^{-5} \right)
\end{equation}
Each merger tree is thereby reduced to a sequence of $x$ values. GIDGET uses this sequence as an input-- starting from any redshift zero halo mass $M_{h,0}$, the accretion rate is computed as $dM_h/dt |_{\mathrm{avg}} 10^x$.  The halo mass $M_h$ is integrated backwards according to this accretion history. 

This methodology relies on the distribution of the logarithm of accretion rates about the median being close to constant across time and halo mass, with only the center of the distribution changing. This approximation is reasonable in the dark matter simulations themselves \citep{neistein_constructing_2008, neistein_universal_2010, rodriguez-puebla_is_2016}, and may be the source of scatter in the star-forming main sequence \citep{forbes_origin_2014, rodriguez-puebla_is_2016}, the mass-metallicity relation, and the anti-correlation in metallicity and star formation rate at fixed mass \citep{forbes_origin_2014}.

Another feature of Equation \ref{eq:randomfactors} is the requirement that $x>-5$. Many of the halo mass assembly histories in Bolshoi contain at least some period of time wherein the halo loses mass. This is the result of a variety of physical effects, as well as the difficulty of correctly assigning N-body particles to halos at each snapshot in the simulation \citep{behroozi_major_2015-1, lee_properties_2016}. Even if the effect is physical, e.g. stripping of a halo by a close encounter, it is unlikely that the galaxy itself will be strongly affected by the halo's mass loss given the difference in physical scale. We therefore simply assume that the accretion rate is essentially zero during this time. There is a danger that some of the subsequent accretion recorded in the accretion history will be spurious (e.g. the halo re-acquiring particles that should have been identified as part of the halo the whole time), but there is no way to distinguish these scenarios from the merger trees alone.

Along with the sequence of $x$ values recorded from the Bolshoi merger trees, we can identify halos that disappear from one snapshot to the next in the tree. For each, again employing the argument that the growth of structure is roughly self-similar, we record the ratio of the mass of the disappearing halo to the mass of the main progenitor halo at that redshift. For simplicity, at each redshift we record the mass ratio of the top two mergers. For the vast majority of halos at a given redshift, there are no mergers between two Bolshoi snapshots, and of the mergers, most correspond to halos near the resolution limit of Bolshoi. When GIDGET reconstructs the mass accretion history starting from an arbitrary new initial halo mass, each merger is re-dimensionalized by multiplying the recorded mass ratio by the halo mass in the reconstructed halo mass history at the same redshift where the merger occurred in Bolshoi. The stellar mass of the merging galaxy is then inferred by employing the stellar-mass halo-mass relation from \citet{moster_constraints_2010}. This stellar mass is tracked as a separate component $M_{*,\mathrm{halo}}$, and has no direct effect on the simulation.

\subsubsection{Radial Distribution of the Accretion Rate}
\label{sec:radialAccretion}
Another consideration regarding accretion is its radial distribution $\dot{\Sigma}_\mathrm{cos}$ in the galaxy. A variety of approaches have been used to address this question. A seminal paper on this subject is \citet{mo_formation_1998}, which assumes the development of an exponential disc with a specific angular momentum (sAM) equal to a fixed fraction of the dark matter sAM. A similar approach accounting for adiabatic contraction and evolution of dark matter internal structure with time was developed by \citet{somerville_explanation_2008} and used in subsequent semi-analytic modelling \citep[e.g.][]{somerville_semi-analytic_2008, popping_sub-mm_2016}. Note that these approaches do not specify $\dot{\Sigma}_\mathrm{cos}$ per se, but rather the final distribution of baryons in the disc. A more detailed method still relying on properties of the dark matter is to estimate the sAM profile of the halo, and the rate at which gas at a particular sAM in the halo cools. That gas is then added to a computational grid discretized in sAM \citep{ van_den_bosch_impact_2002, dutton_revised_2007}.

The fundamental assumption of the approaches above is that gas closely follows dark matter, not just in terms of its density distribution, but also in terms of its angular momentum. Cosmological hydrodynamic simulations show that this is not the case. Rather, gas inflowing onto a galaxy can have comparatively large sAM \citep{danovich_four_2015} as the result of inflow along cold filaments \citep[e.g.][]{dekel_cold_2009}, though the survival of these streams is controversial \citep{nelson_impact_2015, mandelker_instability_2016}. In the picture presented by \citet{danovich_four_2015}, the gas does eventually reach a sAM distribution in line with expectations from simpler considerations \citep{mo_formation_1998}, and suggested by the observations \citep{burkert_angular_2016}, but this may be a coincidence. Indeed, in simulations it appears that, while the dark matter and baryonic sAM content of a population of galaxies is quite similar in magnitude, on a halo-to-halo basis there is little correlation between the two \citep{obreja_NIHAO_2016}.

Given the large theoretical uncertainties, we adopt essentially the simplest assumption, which is that gas accretes in an exponential surface density profile with a scale length equal to a fixed fraction of the instantaneous Virial radius, i.e. $\dot{\Sigma}_\mathrm{cos} \propto \exp( -r/(\alpha_R R_\mathrm{Vir}) )$ This is similar to our approach in \citet{forbes_balance_2014}, and in the semi-analytic work of \citet{fu_star_2013}.

\subsubsection{Quenching}
\label{sec:quenching}
A number of galaxy properties are observed to be bimodally distributed, with disc-like / blue / star-forming galaxies separated from spheroidal / red / quenched galaxies, and relatively few galaxies in between. Although it is possible for low-mass galaxies to fall into the quenched category, such galaxies are found almost exclusively in dense environments. Galaxies appear to undergo a separate quenching process when they grow to a certain halo mass, so that mass and environment can each separately cause quenching \citep{peng_mass_2010, peng_mass_2012}. For the purposes of our model, we are only following galaxies that are the main progenitors of $z=0$ galaxies, and we include no information about each galaxy's position in the universe relative to any other. We are therefore almost exclusively concerned with mass quenching, rather than environment quenching.

The quenching of high-mass galaxies likely requires feedback from supermassive black holes. Exact implementations from different simulations and models vary immensely, and thus far we have not explicitly implemented a model for the growth of black holes in our model. Instead we employ a simple halo quenching model, in which galaxies that reach a certain halo mass suffer severely reduced baryonic accretion \citep{dekel_galaxy_2006, dekel_gravitational_2008}. This has the advantage of being able to capture the turnover in the stellar mass halo mass relation, and account for much of the bimodality observed in galaxies, but is unable to capture the occasional rejuvenation (the transition of a galaxy from red back to blue) of galaxies as suggested both observationally \citep{pandya_nature_2016-1} and theoretically \citep{pontzen_how_2016}. Since our work is mainly focused on the internal properties of spiral and dwarf galaxies, which are not quenched, we do not attempt to implement a more realistic model that could capture these effects.

To implement this form of mass quenching, we adopt the following simple model. The baryonic accretion efficiency specified by equation \ref{eq:acceff} is reduced by a further factor $\epsilon_\mathrm{reduce}$,
\begin{equation}
\label{eq:accreduce}
\epsilon_\mathrm{reduce} = \begin{cases}
1 & M_h <M_{q} \\
\sqrt{\epsilon_\mathrm{q}} & z>1.5+\log_{10}( M_h/ M_{q})\ \mathrm{and}\ M_h>M_{q} \\
\epsilon_\mathrm{q}& z<1.5+\log_{10}( M_h/ M_{q})\ \mathrm{and}\ M_h>M_{q} \\
\end{cases}
\end{equation}
These divisions roughly correspond to those proposed in \citet{dekel_gravitational_2008}. In the absence of a stable accretion shock, gas is able to cool efficiently and accrete onto the galaxy, so the accretion efficiency is not reduced. In the regime where hot halos exist but gas may accrete via narrow cold streams, the efficiency is reduced but only moderately. When the width of accreting streams approaches the virial radius, the galaxy is fully quenched. One could imagine many variations on this scheme, but the exact choice is not obvious, so for now we use this simple prescription.

\subsection{Star Formation}
Stars form according to a combination of ideas outlined in \citet{krumholz_star_2012} and  \citet{krumholz_star_2013}. We assume that the star formation surface density follows
\begin{equation}
\label{eq:colsfr}
\dot{\Sigma}_* = \epsilon_\mathrm{ff} f_{\mathrm{H}_2} \Sigma / t_\mathrm{ff}.
\end{equation}
The efficiency per free fall time $\epsilon_\mathrm{ff}$ is a free parameter of order 1\%, as estimated observationally \citep{krumholz_slow_2007, krumholz_universal_2012, salim_universal_2015, usero_variations_2015, vutisalchavakul_star_2016, heyer_rate_2016, leroy_cloud-scale_2017, onus_numerical_2018} and understood theoretically as a consequence of turbulence in self-gravitating gas \citep[e.g.][]{krumholz_general_2005, federrath_star_2012, padoan_simple_2012}. Finally, the freefall timescale $t_\mathrm{ff}$ can be estimated as the minimum of the freefall time within a single molecular cloud, and the disc crossing time in the case that the ISM is globally gravitationally unstable \citep{krumholz_universal_2012}. That is,
\begin{eqnarray}
t_\mathrm{ff}& =& \min( t_\mathrm{GMC}, t_\mathrm{Toomre} ) \\
t_\mathrm{GMC} &=& \frac{\pi^{1/4}}{\sqrt{8}} \frac{\sigma}{G(\Sigma_\mathrm{GMC}^3\Sigma)^{1/4}} \nonumber \\
t_\mathrm{Toomre} &=& \sqrt{\frac{3 \pi^2 Q_g^2}{64(\beta+1)}}\sqrt{1+\frac{\Sigma_* \sigma}{\Sigma \sigma_\mathrm{zz}}}\frac{r}{v_\phi} \nonumber
\end{eqnarray}
where $\Sigma_\mathrm{GMC} \approx 85 M_\odot/\mathrm{pc}^2$ is a typical GMC surface density \citep{bolatto_resolved_2008}, and as in Equation \ref{eq:netheating}, the second factor in $t_\mathrm{Toomre}$ is a correction for the finite thickness of the disc.

In the past we have estimated the fraction of hydrogen mass in a molecular phase, $f_{\mathrm{H}_2}$, according to \citet{krumholz_atomic--molecular_2008, krumholz_atomic--molecular_2009, krumholz_star_2009}, but this prescription has difficulty reproducing the small but non-zero molecular gas fraction at low column densities. By combining aspects of \citet{krumholz_star_2009} and \citet{ostriker_regulation_2010}, \citet{krumholz_star_2013} developed a model to self-consistently estimate the star formation rate, and molecular hydrogen fraction even in cases where \citet{krumholz_star_2009} broke down. 

The \citet{krumholz_star_2013} model is iterative. Given a fixed total gas surface density $\Sigma$, gas phase metallicity relative to solar $Z'$, density of collisionless matter $\rho_\mathrm{sd}$ (including stars, stellar remnants, and dark matter), and $t_\mathrm{ff}$ as defined above, the model solves for a value of $\dot{\Sigma}_*$ that produces a self-consistent molecular fraction $f_{\mathrm{H}_2}$, interstellar radiation field $G_0'$, and particle density of the cold neutral medium $n_\mathrm{CNM}$. 

The molecular fraction is taken to be
\begin{equation}
f_{\mathrm{H}_2} = 
\begin{cases}
1 - (3/4) s/ (1+0.25 s) & \mathrm{if}\ s<2 \\
0 & \mathrm{if}\ s\ge 2 
\end{cases}
\end{equation}
where
\begin{eqnarray}
s &=& \frac{\ln \left(1+ 0.6 \chi + 0.01 \chi^2 \right)}{0.6 \tau_c} \\
\tau_c &=& 0.066 f_c Z' \Sigma_0
\end{eqnarray}
These equations are the same in \citet{krumholz_star_2009} and \citet{krumholz_star_2013}; here $f_c$ is the clumping factor, which should be $\sim 5$ if these equations are applied on $\ga 1$ kpc scales, or $\sim 1$ on smaller scales. The value of $\tau_c$ is normalized for a column density of $\Sigma_0=\Sigma/1 M_\odot \mathrm{pc}^2$. To evaluate this equation we also need to specify
\begin{equation}
\chi = 7.2 G_0' / n_1
\end{equation}
where $n_1$ is the density of the CNM $n_\mathrm{CNM}$ in units of $10\ \mathrm{cm}^{-3}$. The interstellar radiation field is assumed to scale with the star formation surface density relative to the solar neighborhood value,
\begin{equation}
G_0' = \frac{\dot{\Sigma}_\mathrm{SF}}{2.5 \times 10^{-3} M_\odot \mathrm{pc}^2 \mathrm{Myr}^{-1}}
\end{equation}
The final piece of the model, and the key difference between \citet{krumholz_star_2009} and \citet{krumholz_star_2013}, is that $n_\mathrm{CNM} = \max ( n_\mathrm{CNM,2p}, n_\mathrm{CNM,hydro})$, where
\begin{equation}
n_\mathrm{CNM,2p} = 23 G_0' \left( \frac{1+3.1 Z'^{0.365}}{4.1}  \right)^{-1}
\end{equation}
and
\begin{eqnarray}
n_\mathrm{CNM, hydro} = \frac{\pi G \Sigma_{\mathrm{HI}}^2}{4.4\alpha k_B T_\mathrm{CNM,max}} \bigg\{1 + 2 R_{\mathrm{H}_2} +   \nonumber \\
\bigg[(1+2 R_{\mathrm{H}_2})^2 + \frac{32 \zeta_d \alpha \tilde{f}_w c_w^2 \rho_\mathrm{sd}}{\pi G \Sigma_\mathrm{HI}^2} \bigg]^{1/2} \bigg\} 
\end{eqnarray}
Here  $R_{\mathrm{H}_2} = \Sigma_{\mathrm{H}_2}/\Sigma_{\mathrm{HI}}$, and we are using $\alpha \sim 3$, the ratio of mid-plane pressure in a galaxy to thermal pressure, $T_\mathrm{CNM,max} \approx 243$, the maximum temperature for the CNM \citep{wolfire_neutral_2003}, a correction factor $\zeta_d \approx 0.33$ depending on the shape of the gas density profile, the sound speed of the warm neutral medium $c_w\approx 8 \mathrm{km}/\mathrm{s}$, and the mass-weighted thermal velocity dispersion divided by the $c_w^2$, $\tilde{f}_w \approx 0.5$ \citep{ostriker_regulation_2010}. We adopt the approximated values quoted here for all galaxies at all times. 

\subsection{Feedback and Metals}
\label{sec:metals}
The formation of stars leads to several forms of feedback on the ISM. We make some crude approximations in our model given our lack of spatial information and resolution. In every radial cell, we take the surface density of gas ejected by feedback to be proportional to the local star formation rate surface density. Following \citet{creasey_how_2013}, the proportionality constant $\mu = \dot{\Sigma}_\mathrm{out}/\dot{\Sigma}_*$, known as the mass loading factor, is taken to have a power law dependence on the local surface density and the ratio of gas density to total density within the midplane, i.e. the gas fraction in the terminology of \citet{creasey_how_2013},
\begin{equation}
\mu = \mu_0 \left(\frac{\Sigma}{10\ M_\odot\ \mathrm{pc}^{-2}}\right)^{\alpha_\Sigma} \left( \frac{\Sigma/0.1}{\Sigma+ \rho_{sd} H}\right)^{\alpha_f} \left(\frac{M_h}{10^{12} M_\odot}\right)^{\alpha_M}.
\label{eq:mlf}
\end{equation}
We additionally include an explicit dependence on halo mass, to allow the model to capture the vastly different velocity scales and potential well depths of different galaxies. As in the previous subsection, $\rho_\mathrm{sd}$ is the density of stars and dark matter in the disc mid-plane, and $H$ is the local gas scale height.

The metal content of galactic winds is extremely difficult to observe directly, but can be a large contributor to the metal budget of a galaxy \citep{peeples_budget_2013}. A natural assumption to make is that at the radius where mass is being ejected, the metallicity of this wind material will be the same as the  metallicity of the local ISM. This is plausible, but assumes that the metal-enriched ejecta of supernovae and stellar winds (each of which may contribute appreciably to the ejection of gas from the galaxy) are well-mixed with ambient gas and not preferentially incorporated into the galactic wind. We relax that assumption by introducing the parameter $\xi$, defined so that the metallicity of the galactic wind is 
\begin{equation}
\mathbf{Z}_w = \mathbf{Z} + \xi \frac{\mathbf{y} }{\max( \mu, 1-f_{R,\mathrm{inst}})} 
\label{eq:Zw}
\end{equation}
Note that $\mathbf{Z}$ and the associated quantities in this equation are in general vectors representing an arbitrary number of metal fields. When $\xi=0$, the material being ejected from the galaxy has a metallicity equal to that of the ambient ISM. When $\xi=1$, every newly-produced gram of metals is ejected from the galaxy. The yield, $y$, is the metal mass returned to the ISM per mass of stars formed. As in section \ref{sec:hydro}, $f_{R,inst}$ is the fraction of mass in newly-formed stars that remains in stellar remnants as opposed to being quickly returned to the ISM via supernovae. 

As in \citet{forbes_balance_2014}, metals are added according to the instantaneous recycling approximation, advected in flows within the disc, accreted along with infalling material, and diffused within the disc.
\begin{eqnarray}
\frac{\partial (\Sigma \mathbf{Z})}{\partial t} & = &\frac{1}{2\pi r} \frac{\partial (\dot{M} \mathbf{Z})}{\partial r} + \mathbf{Z}_\mathrm{acc}\dot{\Sigma}_\mathrm{cos} + (\mathbf{y}  \\
& &  - f_{R,\mathrm{inst}} \mathbf{Z} - \mu \mathbf{Z}_w) \dot{\Sigma}_\mathrm{SF} + \frac{\partial}{\partial r} \kappa_Z \frac{\partial (\Sigma \mathbf{Z})}{\partial r} + \mathbf{y}_{Ia}\mathcal{R}_{Ia} \nonumber
\label{eq:metalhydro}
\end{eqnarray}
in analogy to the continuity equation for gas, Equation \ref{eq:dcoldt}. The value of the diffusion coefficient $\kappa_Z$ is taken from \citet{yang_thermal-instability-driven_2012} as in \citet{forbes_balance_2014},
\begin{equation}
\label{eq:kappaz}
\kappa_Z = \min\left( 3.32 \times 10^{-3} k_Z  \frac{\kappa \sigma^4}{G^2 \Sigma^2},\  r\sigma \right)
\end{equation}
where $\kappa$ is the epicyclic frequency, not to be confused with the diffusion coefficient $\kappa_Z$. We include a free parameter $k_Z$ for flexibility, and we require that the diffusion coefficient not exceed $r \sigma$, the product of the largest size scale and velocity scale we would expect to be relevant for turbulent mixing. In practice we end up reducing the allowable values of $k_Z$ to of order a few percent to prevent spurious rapid diffusion in the outer parts of disks where $\sigma^4/\Sigma^2$ can become quite large.

Rather than a single metal field, iron-peak and $\alpha$ elements are tracked separately following \citet{kim_agora_2014}. We therefore treat each of $\mathbf{Z}$, $\mathbf{y}$, $\mathbf{y}_{Ia}$, $\mathbf{Z}_w$ and $\mathbf{Z}_\mathrm{acc}$ as a 2-component vector corresponding to $\alpha$ and Fe. The two components of the yield $\mathbf{y}$ corresponding to the quantity of metals formed per unit mass of star formation in type II supernovae, and $\mathbf{y}_{Ia}$, the mass of metals produced per type Ia supernova are of course different. The values of the yields are included in Table \ref{tab:priors}, and are obtained by multiplying the oxygen or iron yield, per mass of stars formed in the case of Type II supernovae \citep{woosley_nucleosynthesis_2007}, or per Type Ia event \citep[the W7 model of][]{iwamoto_nucleosynthesis_1999}, by the ratio of $\alpha$ elements to oxygen or iron-group elements to iron in the sun \citep{asplund_chemical_2009}. The rate of type Ia supernovae per unit area in the disc is taken to be
\begin{equation}
\mathcal{R}_{Ia} = 2.0 \times 10^{-3} M_\odot^{-1} \sum_i \frac{\Sigma_{*,i}}{1 - f_{ml}(T_i)} \frac{1/\ln(100)}{T_i} \mathcal{I}_{0.1<T_i<10} 
\end{equation} 
The overall normalization is the number of type Ia supernovae per unit stellar mass formed. Our value of $2\times 10^{-3}$ is somewhat larger than that inferred by \citet{maoz_delay-time_2012} observationally, but it has been adjusted by hand to better match the $\alpha/\mathrm{Fe}$ ratio in our simulations. The sum over $i$ stellar populations is a numerical approximation to the true continuous age distribution. The factor of $(1-f_{ml})$ is a matter of accounting; $\Sigma_{*,i}$ is the instantaneous stellar mass surface density, as opposed to the surface density of stars at the time of formation. The difference between the two is this factor assuming a large enough number of populations is used that each may be considered approximately mono-age. This is not a stringent requirement because $f_{ml}$ is only logarithmically dependent on age, with a characteristic timescale far shorter than the 100 Myr lower cutoff described by the indicator function $\mathcal{I}_{0.1<T_i<10}$, which is $1$ when $T_i$ is between 0.1 and 10 Gyr, and 0 otherwise. Finally the factor of $1/\ln(100)$ normalizes the $1/T$ term such that $(1/\ln(100)) \int T^{-1} \mathcal{I}_{0.1<T<10}\ dt =1$. 

A major difference between \citet{forbes_balance_2014} and our prescription here is the inclusion of a galactic fountain term, which allows metals ejected in the galactic wind to be re-incorporated to the accretion flow, and thereby mixed throughout the galaxy, as occurs in cosmological simulations \citep[e.g.][]{oppenheimer_mass_2008, ma_origin_2016}. Therefore instead of $Z_\mathrm{acc} = Z_\mathrm{IGM}$, assumed constant over the course of a simulation, we set
\begin{equation}
Z_\mathrm{acc} = Z_\mathrm{IGM} + \xi_\mathrm{acc} \int 2\pi r \mu \dot{\Sigma}_\mathrm{SF} Z_w dr / \dot{M}_\mathrm{ext}
\label{eq:xiacc}
\end{equation}
The mixing parameter $\xi_\mathrm{acc}$ must be between 0 and 1, corresponding to no mixing of outflows and inflows or total re-accretion of all ejected metals respectively.

\subsection{Initial Conditions}
\label{sec:ic}
Galaxies are initiated at $z=4$ as discs which are exponential in both gas and stars, though the two are allowed to have differing scale lengths. The metallicities are taken to be uniform in radius, and identical to the metallicity of accreting gas prior to mixing with outflowing metals, i.e. $Z_\mathrm{IGM}$. The velocity dispersion of both gas and stars are set such that the galaxy is instantaneously gravitationally stable, i.e. $Q_*>Q_\mathrm{lim}$ and $Q>Q_f$. 

To set the absolute value of the metallicity, stellar mass, and gas fraction we turn directly to observational constraints. The halo mass at the start of the simulation is pre-ordained by the particular accretion history that has been chosen, integrated backwards from the model's given $M_{h,0}$. Given the halo mass, stellar masses are set according to the results of \citet{moster_galactic_2013} at $z=4$ at all masses (though of course this is an extrapolation at low masses). 
\begin{equation}
\label{eq:smhm}
\frac{M_*}{M_h} = 2 N_z \left( \left(\frac{M_h}{M_1} \right)^{-\beta_z-\Delta\beta} + \left( \frac{M_h}{M_1} \right)^{\gamma_z} \right)^{-1}.
\end{equation}
Each parameter, namely $N_z$, $M_1$, $\beta_z$, and $\gamma_z$ is set according to the best-fit values from \citet{moster_galactic_2013}. The low-mass powerlaw slope of this relation, set by $\beta_z$ is given an unknown constant offset $\Delta\beta$. 

The gas mass is set following the fit to observational data given in \citet{hopkins_dissipation_2009}. Given a stellar mass, we compute the gas mass and multiply it by a factor $\chi_{f_g}$,
\begin{equation}
\label{eq:chifg}
M_g = \chi_{f_g} M_*/(1/f_{g,0} -1)
\end{equation}
where at $z=4$ we set the initial guess for the gas fraction $f_{g,0}$ following \citet{hopkins_dissipation_2009}
\begin{equation}
f_{g,0} = \frac{(1 - \tau_4 (1-f_0^{1.5}))^{-2/3}}{1+(M_*/10^{9.15} M_\odot)^{0.4}} 
\end{equation}
and $\tau_4 = 12.27/(12.27+1.60)$ is the fractional lookback time at $z=4$ compared to the age of the universe. Since this is an extrapolation from data which is itself uncertain, we treat this as a rough guess and  assign $\chi_{f_g}$ a large prior range.

To estimate the initial metallicity and the metallicity of subsequent mass accretion, we begin with the simple powerlaw relation between stellar mass and gas-phase metallicity observed by \citet{lee_extending_2006-2} in the local universe for low-mass galaxies. At the beginning of each simulation, once the initial $z=4$ stellar mass is determined (based on $M_{h,0}$, the accretion history, and a stellar mass-halo mass relation as described above), we assign
\begin{equation}
\label{eq:ZIGM}
\log_{10} Z_\mathrm{IGM}  = -3.05 + \chi_{Z,\mathrm{slope}} \log_{10}\left(\frac{M_*(z=4)}{10^{10} M_\odot}\right) + \log_{10} \chi_{Z,\mathrm{offset}}
\end{equation}
The value of $-3.05$ begins with the corresponding value for \citet{lee_extending_2006-2}. The normalization is then adjusted to convert from units of $12+\log_{10}(O/H)$ to absolute metallicity units, and further reduced by a factor of 20, since we expect the IGM to have an appreciably lower metallicity than $z=0$ galaxies. The normalization and slope are then given broad prior ranges.

\section{Searching for a Good Fit}
\label{sec:search}
The model we have described in the previous sections is essentially a more detailed and flexible semi-analytic model. We include more physics, e.g. viscous gas and stellar transport, feedback and star formation dependent on local disc properties, and no imposition of a particular baryonic profile, but in so doing we have more physics to parameterize. In many situations, as described in the previous section, there is a reasonable estimate of the unknown parameter available from observations or simulations. In other cases, as for feedback where even the basic physical processes responsible for driving galactic winds are still unclear, the parameters are poorly constrained.

Our goal in this section is to identify a set of values for the physical parameters described in the previous section that simultaneously reproduce a wide variety of observational data. A natural approach would be to operate within a Bayesian framework, inferring the values of the physical parameters via a MCMC or similar method, where the likelihood function simultaneously compared the results of GIDGET runs to a a diverse set of observational data spanning mass, redshift, and physical characteristics of a galaxy. Unfortunately, this approach done naively is prohibitively expensive due to the runtimes of GIDGET, on the order of 10 minutes, depending on the parameter values. To make the problem computationally feasible, we instead run the MCMC using an emulator, which estimates the results of the GIDGET model given a particular point in parameter space using machine learning, rather than running a new set of models each time the likelihood is evaluated. This corresponds to a speedup of about 6 orders of magnitude in calls to the likelihood function. 

To train the emulator, we must first run GIDGET a reasonably large number of times to provide examples to the machine learning algorithm. Each run provides an example of the map between the inputs (or features), i.e. a point in parameter space, and the outputs (or targets), i.e. observable quantities that can be compared to data, including for example the stellar mass of the galaxy at $z=3$, or the gas-phase metallicity at $z=1$. To generate this training set, we define a distribution over parameter space which will be closely related to (but crucially different from) the prior that will be used in the Bayesian inference step. The training set consists of 41393 model galaxies, which is sufficient to provide accurate predictions for a number of observables with a few caveats. Vigilance against overfitting in certain corners of parameter space is critical, and not all observables can be accurately predicted with this sample, at least with the algorithms we have tried. Nonetheless the emulator's speed allows us to run an MCMC to convergence, producing as many draws of the parameters from the posterior distribution as we would like. These parameters can then be used as inputs to a set of full GIDGET models, allowing us to first verify that the emulator+MCMC have done a good job fitting the data that was included in the likelihood function, and second to make predictions for many other quantities which were not included in the fit. Each step in this process is described in more detail in this section. 

\subsection{Training Set and Priors}
\label{sec:priors}

\begin{table*}
\caption{Parameters controlling individual runs of the GIDGET model.}
\begin{tabular}{lllll}
& Parameter & Defined & Varied & Training Set Distribution\footnote{testing12345 678} \\
\hline
\hline
{\bf Initial Conditions} & & & & \\
Halo mass at $z=0$ & $M_{h,0}$ & & Varied & $log-\mathcal{U}(10^{11} M_\odot, 10^{12} M_\odot)$  \\
Reduction in initial stellar scale radius & $\alpha_{r,*,0}$ & Section \ref{sec:ic} & Varied & $log-\mathcal{N}(2, 0.3\ \mathrm{dex})$ \\
Reduction in initial gas scale radius & $\alpha_{r,g,0}$ & Section \ref{sec:ic} & Varied & $log-\mathcal{N}(2, 0.3\ \mathrm{dex})$ \\
Initial metallicity relative to a fiducial guess & $\chi_{Z,\mathrm{offset}}$ & Equation \eqref{eq:ZIGM} & Varied & $log-\mathcal{N}(1, 1\ \mathrm{dex})$\\
Initial metallicity relative to a fiducial guess & $\chi_{Z,\mathrm{slope}}$ & Equation \eqref{eq:ZIGM} & Varied & $\mathcal{N}(0.3, 0.2)$\\
Initial Gas Mass relative to a fiducial guess & $\chi_{f_{g,0}}$ & Equation \eqref{eq:chifg} &Varied  & $log-\mathcal{N}(2, 0.3\ \mathrm{dex})$ \\
$\log_{10}$ offset in concentration-mass relation & $\alpha_\mathrm{con}$ & Equation \eqref{eq:alphacon} & Varied & $\mathcal{N}(0, 0.3)$ \\
Modification to low-mass SMHM slope & $\Delta\beta$ & Equation \eqref{eq:smhm} & Varied & $\mathcal{N}(0,0.3)$\\
\hline
{\bf Accretion} & & & & \\
Maximum accretion efficiency & $\epsilon_\mathrm{max}$ & Equation \eqref{eq:acceff} & Varied & $\mathcal{U}(0,1)$ \\
Quenching mass (M$_\odot$) & $M_\mathrm{q}$ & Equation \eqref{eq:accreduce} & Varied & $log-\mathcal{N}(10^{12},0.48\ \mathrm{dex})$ \\
Quenching efficiency & $\epsilon_\mathrm{q}$ & Equation \eqref{eq:accreduce} & Varied & $log-\mathcal{N}(10^{-3},1\ \mathrm{dex})$ \\
Proportionality constant of accretion scale radius & $\alpha_r$ & Section \ref{sec:radialAccretion} & Varied & $log-\mathcal{N}(0.141, 0.48\ \mathrm{dex})$ \\
\hline
{\bf Outflows} & & & & \\
Scaling of mass loading factor with $\Sigma$ & $\alpha_\Sigma$ & Equation \eqref{eq:mlf} & Varied & $\mathcal{N}(0, 1)$ \\
Scaling of mass loading factor with $f_g$ & $\alpha_f$ & Equation \eqref{eq:mlf} & Varied & $\mathcal{N}(0, 2)$ \\
Scaling of mass loading factor with $M_h$ & $\alpha_{M_h}$ & Equation \eqref{eq:mlf} & Varied & $\mathcal{N}(-1, 1)$ \\
Normalization of mass loading factor & $\mu_0$ & Equation \eqref{eq:mlf} & Varied & $log-\mathcal{N}(0.01, 1\ \mathrm{dex})$ \\
Mixing of SN ejecta with stellar winds & $\xi$ & Equation \eqref{eq:Zw} &Varied & Beta(1,2) \\
Mixing of wind material with inflows & $\xi_\mathrm{acc}$ & Equation \eqref{eq:xiacc} & Varied & $\mathcal{U}(0,1)$\\
\hline
{\bf Star Formation} & & & & \\
Efficiency per freefall time & $\epsilon_\mathrm{ff}$ & Equation \eqref{eq:colsfr} & Varied & $log-\mathcal{N}(0.01,  0.3\ \mathrm{dex})$ \\
Adjustment to asymptotic SN momentum injection & $\chi_\mathrm{inj}$ &  Equation \eqref{eq:netheating} & Varied & $log-\mathcal{N}(1,  0.3\ \mathrm{dex})$ \\
Short timescale remnant fraction & $f_{R,\mathrm{inst}}$ & Section \ref{sec:hydro}  & Fixed & 0.77 \\
Asymptotic remnant fraction & $f_{R,\mathrm{asym}}$ & Section \ref{sec:rot} & Fixed & 0.503\\
Yield of $\alpha$-elements from type II SNe & $y_{\alpha}$ & Section \ref{sec:metals} & Fixed & 0.0278\\
Yield of iron peak-elements from type II SNe & $y_\mathrm{Fe}$ & Section \ref{sec:metals} & Fixed & 0.00117 \\
Yield of $\alpha$-elements from type Ia SNe & $y_{Ia,\alpha}$ & Section \ref{sec:metals} & Fixed & 0.2926 \\
Yield of iron peak-elements from type Ia SNe & $y_{Ia,\mathrm{Fe}}$ & Section \ref{sec:metals} & Fixed & 0.6678 \\
\hline
{\bf In-disc Transport} & & & & \\
Dissipation rate per disc height crossing time & $\eta$ & Equation \eqref{eq:netheating} & Varied & $log-\mathcal{N}(1.5, 0.3\ \mathrm{dex})$ \\
The thermal velocity dispersion of the WNM & $\sigma_\mathrm{sf}$ & Equation \eqref{eq:netheating} & Fixed & 7.6 km/s \\
Threshold value of $Q$ & $Q_f$ & Equation \eqref{eq:dQdt} & Varied & $log-\mathcal{N}(1.5, 0.3\ \mathrm{dex})$\\
Additional Shakura-Sunyaev $\alpha$ viscosity & $\alpha_\mathrm{MRI}$ & Equation \eqref{eq:alphaMRI} & Varied & $log-\mathcal{N}(0.05, 0.3\ \mathrm{dex})$\\
In-disc metal mixing rate & $k_Z$ & Equation \eqref{eq:kappaz} & Varied & $log-\mathcal{N}(0.025, 0.3\ \mathrm{dex})$\\
\hline
{\bf Numerics} & & & & \\
Number of cells & $n_x$  & Section \ref{sec:hydro} &  Fixed & 256 \\
$k$ cutoff scale to suppress $v_\phi$ oscillations & $k_\mathrm{lim}$ &  Equations \eqref{eq:fft} and \eqref{eq:ifft}   & Fixed & 10 \\
Power to which to raise exponential cutoff & $n_\mathrm{lim}$ & Equations \eqref{eq:fft} and \eqref{eq:ifft} & Fixed & 2 \\
Inner boundary of computational domain & $r_\mathrm{min}$ & Section \ref{sec:hydro} & Fixed & $10^{-3}$ R \\
Outer boundary of computational domain & $r_\mathrm{max}$ & Section \ref{sec:hydro} & Fixed & $119 (M/10^{11} M_\odot)^{1/4} (\alpha_r/0.1)\ \mathrm{kpc}$ \\
Minimum stellar velocity dispersion & $\sigma_{*,\mathrm{min}}$ & Section \ref{sec:hydro} & Fixed & $10\ \mathrm{km}\ \mathrm{s}^{-1}$\\
Starting redshift & $z_\mathrm{start}$ & & Fixed & 4 \\
\hline
\end{tabular} \par
\label{tab:priors}
\bigskip
$^2$ The Prior distributions are either normal, denoted $\mathcal{N}$(mean, standard deviation), log-normal, denoted $log-\mathcal{N}$(median, standard deviation in $\log_{10}$ of the variable), uniform, denoted $\mathcal{U}$(minimum, maximum), log-uniform, denoted $log-\mathcal{U}$(minimum,maximum), or Beta (whose PDF is $\propto x^{\alpha-1}(1-x)^{\beta-1}$), denoted $Beta(\alpha,\beta)$.
\end{table*}

The first step in this process is to define the parameter space through which we will search. Most of these parameters have been defined in the previous section. Table \ref{tab:priors} summarizes the parameters used and the distributions from which they are drawn when creating the training set. The joint distribution of these parameters is the product of the distribution on the individual parameters, i.e. each parameter is assumed to be independent. 

Ultimately this distribution will serve as the basis for an informative Bayesian prior, i.e. we are actively making judgements about the acceptable ranges these parameters can have. This is unavoidable when using an emulator because the general-purpose fitting algorithms we use perform poorly outside of the region of parameter space covered by the training set.

To set the values used in Table \ref{tab:priors}, we use log-normal distributions for variables parameterizing uncertain physics but requiring positive values, typically simply asserting factor of 2 or 3 uncertainties. This includes $\epsilon_\mathrm{ff}$, $\eta$, $Q_f$, and $\alpha_r$. The star formation efficiency per freefall time $\epsilon_\mathrm{ff}$ has had its uncertainty estimated empirically in \citet{krumholz_star_2012}. The energy dissipation rate per disc crossing time $\eta$ has been measured in idealized simulations \citep{stone_dissipation_1998, mac_low_kinetic_1998}, but could be somewhat different in a realistic galactic disc \citep[see][for example]{birnboim_compression_2018}. The critical value of $Q$, $Q_f$, is well known to be unity in the isothermal linear regime \citep{toomre_gravitational_1964}, but it is likely somewhat higher in galactic discs with more realistic physics \citep{elmegreen_gravitational_2011, inoue_non-linear_2016}.

The parameters controlling the mass loading factor are substantially more uncertain. The predicted value of the mass loading factor varies substantially \citep{ zahid_census_2012, vogelsberger_physical_2013, muratov_gusty_2015,  schroetter_muse_2016}, and even within a single study, the statistical uncertainty in how the mass loading factor scales is substantial \citep{creasey_how_2013}. As a result, the lognormal scatter in the normalization, $\mu_0$ is much larger than a factor of 2, and the allowed scatter in the scaling exponents is large. A value of $\mu_0$ which gave reasonable fits to the stellar mass halo mass relation was found by hand, and the training set distribution of $\mu_0$ is centered near this value. This central value is substantially smaller than 1 since other factors in the local value of $\mu$ can be quite large, if their powerlaw indices are anything besides zero. One of the most important aspects of the value of $\mu$ is whether or not it is above or below $\sim 1$. When $\mu \la 1$, it plays a minor role in setting the equilibrium values of the column density \citep{forbes_balance_2014} and metallicity \citep{forbes_origin_2014} of the disc, while for $\mu \ga 1$, it may become the dominant contributor \citep{lu_analytical_2015}. By setting $\mu_0$ to a typically small value, the model is given freedom to determine which column densities, gas fractions, or halo masses will have $\mu \ga 1$.

Several parameters are physically constrained to lie between 0 and 1. For most of these we adopt a relatively uninformative distribution, namely the uniform distribution from 0 to 1. These include $\xi_\mathrm{mix}$, the degree to which outflowing metals are re-incorporated in inflows, and the maximum accretion efficiency $\epsilon_\mathrm{max}$. The degree to which supernova ejecta are mixed into the ISM before they participate in the launching of a galactic outflow, $\xi$, is, while still broadly distributed, pushed towards lower values namely $Beta(1,2)$. This reflects the fact that many models assume $\xi=0$, largely because it is the simplest assumption. We also know that $\xi$ must not be too close to unity, since galaxies undoubtedly retain some of the metals they produce. Formally the quenching efficiency $\epsilon_\mathrm{quench}$ must also be between 0 and 1, but we expect $\epsilon_\mathrm{quench} \ll 1$, so we adopt a lognormal distribution centered on a small value.

Given this distribution of physical parameters, we then proceed to draw points from this distribution and run GIDGET models. To do so we also need to specify a halo mass and an accretion history. These quantities are not fit in the inference process, and so will not have a corresponding factor in the prior. For the training set, we simply choose a $z=0$ halo mass log-uniformly distributed between $10^{11}$ and $10^{12} M_\odot$, and a random draw from the Bolshoi accretion histories. The $z=0$ halo mass and a particular average of the set of $x$ values specifying the accretion history are included as features that the machine learning algorithm has access to - basically the purpose of the emulator is to do the job of the full GIDGET model, so the emulator should be provided with as much information as possible.

Each time a point from parameter space is drawn, including a $z=0$ halo mass and accretion history, a single GIDGET model is run. Because the range of parameters is quite large, there can be unexpected combinations of parameters that are pathological and produce models that crash or take too long to run. The scale in parameter space at which these regions arise can be both large and small. That is, models can fail both in well-defined large regions of parameter space that produce unrealistic instabilities, and in otherwise well-behaved regions because for the exact values being simulated, one cell of the simulation happens to develop a sharp feature that limits the timestep to very small values. These model failure regimes have different implications for the emulator that will be discussed in the next section. The 41393 model galaxies quoted as the size of the training set earlier in this section refers to the number of successfully-run models.

\subsection{Emulator}

The training set provides many examples of the map from a set of parameters that we denote $\Theta$ to a set of outputs denoted $\psi$. The values of $\psi$ will be used in the likelihood function discussed in the next subsection. Explicitly, 
\begin{eqnarray}
\label{eq:thetatrain}
\Theta = \{ M_{h,0}, \alpha_r, \alpha_{r,*,0}, \alpha_{r,g,0}, \chi_{f_g,0}, \nonumber \\
 \alpha_\Sigma, \alpha_{f_g}, \mu_0, \alpha_{M_h}, \nonumber \\ 
\chi_{Z,\mathrm{slope}}, \chi_{Z,\mathrm{offset}}, \xi_\mathrm{mix}, \nonumber \\
 \eta, Q_f, \alpha_\mathrm{MRI}, \epsilon_\mathrm{quench}, \epsilon_\mathrm{max}, \alpha_\mathrm{con}, k_Z, \nonumber \\
\xi, \epsilon_\mathrm{ff}, \Delta\beta, M_\mathrm{quench}, \chi_\mathrm{inj}, \langle 10^{x} \rangle_{i=1}^{16}    \}.
\end{eqnarray}
This is a combination of the $z=0$ halo mass, $M_{h,0}$, the physical quantities we would ultimately like to constrain, and a representation of the accretion history denoted $\langle 10^{x} \rangle_{i=1}^{16}$. This is simply shorthand to say that, instead of using the full 1000-element represnetation of the accretion history, we average the values of $10^x$ within constant-sized redshift intervals of $\Delta z = 0.25$, so that the redshift range between $z=0$ and $z=4$ is divided into 16 intervals.

Each time a model is run, we extract and record 200 quantities based on the model output and discard the full model history to avoid storing excessive amounts of data. These 200 quantities are 20 different ``integrated'' quantities, i.e. quantities that involve a sum or average over the entire computational domain, at 4 different redshifts ($z=0,1,2$ and $3$).  We also record 6 quantities at 20 different radii at $z=0$. Since many of these quantities will not enter the likelihood function, and several proved difficult to emulate with our particular training set, we focus on the following 7 integrated quantities, which will be predicted at the 4 different redshifts for a total of 28 targets for the emulator. Explicitly the integrated quantities are
\begin{equation}
\label{eq:targets}
\psi = \{ M_h, M_*, sSFR, \langle Z\rangle_\mathrm{SF}, M_{\mathrm{H}_2}/M_*, r_*, \langle\sigma\rangle_\mathrm{SF}, v_{2.2}\},
\end{equation}
The halo mass $M_h$ has the same meaning it has throughout the paper. The stellar mass $M_*$ refers to instantaneous stellar mass obtained by adding up $\Sigma_*$ times the area of each cell in the computational domain, and adding $M_\mathrm{central}$ (the stellar mass contained within the inner radius of the computational domain as discussed following Equation \eqref{eq:mdotcentral}). In order to compare this to observations based on SED fitting, we will have to adjust this mass to account for stellar mass that has already been returned to the ISM (where we have made this adjustment, we will denote the stellar mass $M_{*,\mathrm{orig}}$. The specific star formation rate is the star formation rate divided by $M_*$, where the star formation rate is obtained by summing $\dot{\Sigma}_*^\mathrm{SF}$ times the area of each annulus, and adding the second term in Equation \eqref{eq:mdotcentral}, which corresponds to short-timescale star formation occurring in the unresolved central part of our galaxy. Both the gas phase metallicity and the gas velocity dispersion are averaged over the star formation rate, meaning they are summed at each radius weighted by $\dot{\Sigma}_*^\mathrm{SF}$ times the area of the cell, then divided by the star formation rate. This is a crude way of imitating observational effects, since both metallicity and velocity dispersion at higher redshifts are estimated using light from star forming regions, so the inferred average values of $\sigma$ and Z in the observations should be light-weighted, or roughly star-formation-weighted. The H$_2$ mass is simply integrated over the computational domain, and $r_*$ refers to the half mass radius of the stars, counting stellar mass in the same way we describe above to determine $M_*$. The circular velocity at 2.2 scale lengths is estimated by evaluating $v_\phi$ in the simulation at $ r_* \cdot 2.2/1.68$, where the factor of $1.68$ is the ratio between the half mass radius and the scale length of an exponential disk.

Ideally we would like to find a mapping from $\Theta$ to $\psi$ that is fast to evaluate and accurate, at least for values of $\Theta$ encompassed by the training set. Ideally we would also like a single map from $\Theta$ to $\psi$, and not many maps from $\Theta$ to each separate component of $\psi$. This has the advantage of better-preserving covariance between different components of $\psi$, and speeding up evaluation of predictions for $\psi$. 

This problem is well-suited to a variety of machine learning regression algorithms, and we have experimented with many of the options available in sklearn \citep{pedregosa_scikit-learn_2012}. Each regression scheme involves a fundamentally different approach, with a corresponding set of hyperparameters. We have found that in general the things that make the largest difference in the performance of the regression are the choice of algorithm, the standardization of the inputs, and excluding hard-to-fit targets from the fit. Particular values of the hyperparameters make relatively little difference (unless extreme values are chosen).

Following their success in \citet{kamdar_machine_2016}, we experimented with different versions of random forests, which make predictions by averaging the results of many individual decision trees. We ultimately moved away from random forests for two reasons: their performance as measured by $R^2$ tended to be a bit lower than other algorithms for this particular problem, and their predictions were piecewise-constant. In other words, within some volume of parameter space, if none of the decision trees happened to predict a different value of a given quantity, the random forest produces the same prediction for every point in that volume. While this is not necessarily a problem if the goal is to maximize accuracy without overfitting the data, it does not work well when those predictions are entering the likelihood function of an MCMC because within these volumes of parameter space, there are no gradients, and the resulting posterior distribution is noticeably pixelated when projected into two dimensions.

In addition to random forests, we experimented with regularized linear regression. These models had several key advantages. First, they performed quite well in terms of $R^2$, particularly when the input and output parameters were scaled to have zero mean and unity variance, and particularly when cross-terms were included in the feature vector. In other words, the algorithm was given not just $\Theta$, but also the product of every pair of quantities in $\Theta$. In addition to their high accuracy, the linear models were relatively insensitive to the inclusion of moderately-difficult targets, and worked extremely well with the MCMC algorithm as a direct result of the fact that all predictions were quadratic in the features; the MCMC would quickly converge to the global maximum without any difficulty. It is possible that by more carefully selecting which cross-terms to include, and potentially including at least a few even higher-order quantities, (e.g. $M_{h,0}^3$), the accuracy could have been improved even further.

In the end, we settled on using neural networks simply because they were the most accurate. High accuracy is absolutely crucial in this problem because the MCMC algorithm is excellent at finding the best fit parameters according to the emulator. If some corner of parameter space where the emulator does a poor job happens to yield a high posterior probability, the MCMC will converge around this point, yielding a solution that does not actually fit the data. To further reduce the the likelihood of this outcome, we combine the predictions of 3 different neural networks by taking the median value of the 3 predictions for each target. In particular the 3 neural networks all have regularization parameters $\alpha=10^{-5}$ and use tanh activation functions, but they have different structures. One has 3 layers of 100 neurons each, one has a 556-neuron first layer with a 111-neuron second layer, and the third network has a 1112-neuron first layer with a 222-neuron second layer. These numbers are multiples (1x and 2x) of the size of the that can in principle learn a given dataset with negligibly small error according to \citet{huang_learning_2003}. This particular combination was one of a few that we tried and seems to perform the best overall on the metrics discussed in Appendix \ref{app:modelcompare} of all the models we tried. We should note though that the largest gain in performance comes from using any decently-sized neural network with $\alpha \la 10^{-2}$, with scaled inputs (see below), and having eliminated any difficult-to-predict components of $\psi$.

In order to make these assessments and evaluate the performance of different models, we follow a rudimentary version of the standard procedure in machine learning: we reserve 10\% of the models produced for the training set for use as a validation set. Each model is trained on the remaining 90\% of the training set. Once the model is trained, the model is used to predict the values of $y$ given the values of $\Theta$ in the validation set. The predicted values of $y$ are then compared to the true values of $y$ in a variety of ways that will be discussed in Appendix \ref{app:modelcompare}.

To make the regression behave more sensibly, we normalize both the features and targets in the following sense. Logarithms are taken of quantities which are amenable to being viewed in log-space - this is a subjective assessment, but in practice it means any quantity except those which can take on negative values, the exponents used in some of the physics prescriptions described in the previous section, and quantities whose priors have appreciable mass over the whole interval from 0 to 1, e.g. $\epsilon_\mathrm{max}$. Additionally, the features are standardized, i.e. subjected to a linear transformation such that their median becomes zero and their distance to the median is reduced by the interquartile range. We note that the transform uses only information from the training set, since we do not want even a small amount of information about the out-of-sample validation set to be used when constructing the fit.

Despite the good performance of the emulator, great care must be taken when using it in the context of an MCMC. As described above and in more detail in Appendix \ref{app:modelcompare}, the performance of each version of the emulator is assessed by comparing the true values of a validation set to the values predicted by the model when applied to the features of that data set. This validation set is just a random subset of the original training set, so no assessment is being made of the model's performance outside of this original training set. Indeed the performance metrics are weighted towards where the training set is densest, where we also expect the regression itself to be the most accurate. Therefore, despite good performance overall, the emulator is likely to be unreliable in regions of parameter space sparsely covered by the training set. These locations are largely in the tails of the distribution used to draw the training set, but they can also arise in regions where GIDGET itself is more likely to fail.

In order to steer the MCMC away from regions where the emulator is likely to be unreliable, we implement the following safeguards. The prior used in the inference process is a narrower version of the distribution used to draw the training set. In particular, every normal or log-normal distribution has its width (in linear- and log-space respectively) reduced by 30\%. In addition to this reduction, we impose a hard boundary on the prior distribution at $\pm 2\sigma$ (again in linear- and log-space respectively). Outside of this range, the prior is identically zero. Additionally the prior on $\xi$, which has a $Beta(1,2)$ in the training set distribution is narrowed to $Beta(1,3)$, concentrating more of the probability mass in at lower values of $\xi$. The priors on $\Delta\beta$ and $\chi_{f_{g,0}}$ were also altered after initial runs of the MCMC to avoid solutions where formally more baryons than the cosmic baryon fraction were included in the initial conditions. In particular the new priors for these variables became $\Delta\beta \sim \mathcal{N}(0, 0.7 \cdot 0.2)$ (where the factor of $0.7$ is the same one that was applied to all normally and log-normally distributed variables), and $\chi_{f_{g,0}} \sim log-\mathcal{N}(1.5, 0.7\cdot 0.11\ \mathrm{dex}) $.

To address regions of parameter space where models may be sparse owing to unforeseen pathological interactions between extreme values of certain parameters, we add a `veto' layer to the prior. That is, we search for a way to identify these regions automatically, and when such a region is identified, the prior in that region is set to zero. We are making the assumption that if GIDGET models consistently fail in that region of parameter space, that region is unlikely to contain the true parameters governing galaxies. An example case arises when $\alpha_{f_g}$ is sufficiently negative; as the gas fraction decreases, the mass loading factor grows, decreasing the gas fraction further. The process runs away until the simulation is ended by passing a lower limit for the timestep. This particular instability likely accounts for most of the predictable model failures encountered when constructing the training set\footnote{The unpredictable failures seem to be associated with the separation between $Q_f$ and $Q_\mathrm{lim}$, the critical $Q$ values for the multi-component disk and the stars alone, respectively. When these values are set too close to each other, it increases the probability that the model will run into a numerical difficulty in which extreme response by the gas is necessary to maintain marginal gravitational instability because the stars by themselves contribute substantially to the 2-component $Q$ if they haven't had sufficient time to heat up.}. 

We explored a few ways to determine veto regions. The main goal was to steer the MCMC clear of regions devoid of successful models while being careful not to throw out valid regions of parameter space. Our training set for this task is similar to the training set used for the regression problem, except this time all $\sim 40000$ successful runs are treated simply as one class, and the $\sim 120000$ models that failed are treated as another class. Simple classification schemes tended not to produce substantial improvements because most of these failures were basically evenly distributed throughout the training set. Instead we used a logistic regression with L1 regularization and a regularization parameter $C=1$. No cross-terms of $\Theta$ were used, just $\Theta$ itself. The logistic regression provides a prediction for the probability that a given model is a failure. We set this threshold to maximize the quantity $F_\beta = (1+\beta^2) \mathcal{P} \mathcal{R} / (\beta^2\mathcal{P} + \mathcal{R})$, which measures the quality of a classification when recall ($\mathcal{R}$) is valued $\beta$ times as much as precision $\mathcal{P}$. We use $\beta = 0.01$ to reflect our desire to avoid throwing out valid regions of parameter space. Because of the small value of $\beta$ and the wide distribution of the failures, this step only ends up excluding about 1\% of the failures, but about 95\% of the models excluded are failures in the validation set.

With the veto region, the narrower parameter space, and the hard limits on how far a parameter is allowed to deviate from its prior, the emulator is ready for use in evaluating the likelihood function.

\subsection{Likelihood function}

\begin{table*}
\caption{Observational relations used to fit the model, or just for visual comparison.}
\label{tab:obs}
\centering
\begin{tabular}{lllll}
Relation & See & Used in fit at redshift$^1$ & Assumed Scatter & Reference(s) \\
\hline
\hline
$M_*$-$M_h$ & Figure \ref{fig:smhm} & 0,1,2,3 & $0.15\ \mathrm{dex}$ & \citet{moster_galactic_2013} \\
& & - & 0.19 & \citet{behroozi_average_2013} \\
& & - & 0.2, 1 & \citet{garrison-kimmel_organized_2017} \\
\hline
sSFR-$M_*$ & Figure \ref{fig:ssfr} & - & - & \citet{brinchmann_physical_2004} \\
 & & - & - & \citet{whitaker_constraining_2014} \\
 & & 0,1,2,3 & 0.34 dex & \citet{lilly_gas-regulation_2013} \\
 & & 0,1,2,3 & 0.28 dex & \citet{speagle_highly_2014-1} \\
\hline
$Z_*$-$M_*$ & Figure \ref{fig:ssfr} & - & 0.17 dex & \citet{gallazzi_ages_2005, kirby_universal_2013} \\
\hline
$Z$-$M_*$ & Figure \ref{fig:ssfr} & 0 & 0.117 dex & \citet{tremonti_origin_2004, lee_extending_2006-2} \\
 & & 1,2,3 & - & \citet{genzel_combined_2015} \\
 \hline
 $M_{\mathrm{H}_2}/M_*$ - $M_*$ & Figure \ref{fig:ssfr}& 0,1,2,3 & - & \citet{genzel_combined_2015} \\
 & & - & - & \citet{saintonge_cold_2011} \\
 \hline
 $M_\mathrm{HI}/M_*$ - $M_*$ & Figure \ref{fig:ssfr} & - & - & \citet{peeples_constraints_2011-1, papastergis_direct_2012} \\
 \hline
 $\Sigma_1$ - $M_*$ & Figure \ref{fig:mre} & - & - & \citet{fang_link_2013} \\
  & & - & - & \citet{barro_structural_2017} \\
 \hline
$r_*$ - $M_*$ & Figure \ref{fig:mre} & 0,1,2,3 & 0.1 dex & \citet{van_der_wel_3d-hst+candels_2014} \\
 & & 0 & - & \citet{baldry_galaxy_2012} \\
\hline
$r_{\mathrm{HI}}$ - $M_\mathrm{HI}$ & Figure \ref{fig:nonmst} & - & - & \citet{broeils_short_1997} \\
\hline
$c_{82}$ - $M_*$ & Figure \ref{fig:mre} & - & 20\% & \citet{dutton_origin_2009} \\
\hline
$v_{\phi,2.2}$ - $M_*$ & Figure \ref{fig:mre} & 0 & 0.058 dex & \citet{miller_assembly_2011, miller_assembly_2012} \\
\hline
$\Sigma/\Sigma_n$ - $r/R_\mathrm{vir}$ & Figure \ref{fig:kravtsovStruct} & - & - & \citet{kravtsov_size-virial_2013} \\
\hline
\end{tabular}\par
\bigskip
$^1$ Dashes in this column (-) indicate that the data is plotted in the referenced figure, but is not used in the likelihood function.
\end{table*}

The regressions discussed in the previous section allow us to quickly evaluate, given a point in the space of model parameters, any target quantity predicted by the regression model. The next step is to construct a likelihood function that compares these fitted quantities to a variety of observational data about galaxies. Table \ref{tab:obs} shows a summary of the observational data that we will employ later (in Figures \ref{fig:smhm}, \ref{fig:ssfr}, and \ref{fig:mre}), a subset of which are used in the likelihood function.

We construct the likelihood function $\mathcal{L}$, formally the joint probability distribution function of the dataset $\mathcal{D}$ given a particular set of model parameters $\Theta^\mathrm{inf}$, as follows. Note that the model parameters we are inferring here are not quite the same as the set of features that are used as inputs to the emulator (denoted $\Theta$ in the previous section). In particular, $\Theta^\mathrm{inf}$ does not include the halo mass or accretion history: the values of those quantities are not up for constraint, but rather the likelihood function has a fixed set of halo masses and accretion histories it uses throughout the entire inference process. We also include several parameters in $\Theta^\mathrm{inf}$ to help account for additional systematic errors. First, we allow the errorbars on certain relations to be increased by a factor $f_\sigma$ with a prior given by a Pareto distribution\footnote{The probability density function for a Pareto-distributed variable $x$ with shape parameter $\alpha$ is $\alpha/x^{\alpha+1}$ for $x>1$, and $0$ otherwise.} with shape parameter $\alpha=3$. In practice, the MCMC always prefers to leave this value at $f_\sigma = 1$ except in the early phases of the MCMC run. We also allow each variable in $\psi$ to be adjusted by a constant factor, since each of the quantities to which we are comparing is itself derived from the data via a model. For now we keep these quantities fixed in mass and redshift. We denote the set of predictions that will be used in the likelihood following these adjustments
\begin{eqnarray}
\tilde{\psi} =& \{ M_h(z), M_*(z) 10^{\delta M} , sSFR(z) 10^{\delta SFR - \delta M}  , \\
 &\langle Z\rangle_\mathrm{SFR}(z) 10^{\delta Z_g}, M_{\mathrm{H}_2}(z)/M_*(z) 10^{\delta M_{\mathrm{H}_2} - \delta M}, \\
 & r_*(z) 10^{\delta r}, \langle\sigma\rangle_\mathrm{SFR}(z) 10^{\delta\sigma} , v_{2.2}(z) 10^{\delta v}\},
\end{eqnarray}
Each $\delta$ is given a truncated normal prior distribution, with the truncation at 2$\sigma$. Most of these truncated gaussians have widths of 0.1 dex, except $\delta r$, $\delta v$, and $\delta\sigma$, which each have widths of 0.05 dex since these quantities are less reliant on models for their derivation.

At the most basic level, we would like to compare the models to the data assuming something resembling a normal distribution, and as is common practice we will assume each data point is independent allowing us to simply multiply the densities to obtain the joint density. Despite this simplifying assumption, we are still faced with several hurdles - first, the observations are not symmetric about the median. Second, the coverage of the data is highly variable - different observational datasets have coverage over different mass ranges which change as a function of redshift. Third, it is often the case that at a fixed mass, different observational datasets have mutually inconsistent estimates of the observable in question. 

To account for the asymmetry, we use a 1D density function similar to the normal distribution, but with a skewness parameter $\epsilon$ from \citet{mudholkar_epsilonskewnormal_2000} \citep[for an approach similar in spirit see][]{espinoza_limb_2015}. Its canonical form is
\begin{equation}
f_0(x | \epsilon) = (2\pi)^{-1/2} \begin{cases}
\exp\left(-\frac{x^2}{2(1+\epsilon)^2} \right) \ \mathrm{if}\ x<0  \\
\exp\left(-\frac{x^2}{2(1-\epsilon)^2}  \right)\ \mathrm{if}\ x\ge 0
\end{cases}
\end{equation}
for $-1<\epsilon<1$. Adding a shape and location parameter, the full ``epsilon-skew-normal'' PDF is $ESN(x|\mu,\sigma,\epsilon) = \sigma^{-1} f_0( (x-\mu)/\sigma\ |\ \epsilon)$. Given the $16$th, $50$th, and $84$th percentiles of the data at a fixed mass, which we'll denote $q_{16}$, $q_{50}$, and $q_{84}$, (typically these are linearly interpolated values in log space because these quantities are only given at a discrete set of masses), we can compute $\mu$, $\sigma$, and $\epsilon$ such that at that fixed mass the ESN distribution has the same quantiles as the data. The quantile function is given analytically in \citet{mudholkar_epsilonskewnormal_2000}, so it is straightforward to numerically solve for $\epsilon$ given the value of e.g. $(q_{84}-q_{50})/(q_{50}-q_{16})$, which cancels out the shape and location parameters, yielding a pure function of $\epsilon$. Once $\epsilon$ is known, $\sigma$ and $\mu$ can be found analytically. Note that this computation is always done in log space, since the distributions of the observable quantities at a fixed mass are roughly log-normal. 

To address the second issue, namely uneven data coverage, we considered simply not counting model predictions that lie outside the mass range covered by the data. This has the disadvantage of introducing a sharp discontinuity in the likelihood function, and incentivizing a pile-up of galaxies on one side or another of this cutoff. Although the sampler is not totally free to adjust the stellar masses to achieve this since it must also fit the stellar mass-halo mass relation, it may be willing to pay a penalty in that fit to stay on the correct side of the cliff in the likelihood. To avoid discontinuities of this sort, we proceed as follows. Outside of the covered mass range, we linearly extrapolate $q_{50}$ in log-space. For the other quantiles, we take $q_{84}-q_{50}$ and $q_{50}-q_{16}$ at the last point covered by the data, and assert that these values inflate exponentially with log distance from the last mass at which there is data by a factor $\exp(4 (\Delta\log_{10}M)^2)$. In other words, we retain the asymmetry of the closest data point, but increase the size of the error bars. Doing so too quickly risks behavior similar to the sharp discontinuity, while doing so too slowly risks constraining the model with a pure extrapolation.

Finally, to address the issue of multiple conflicting datasets at a fixed mass, we simply set the likelihood to an even mixture of the different datasets. The final likelihood function is then
\begin{eqnarray}
\mathcal{L}(\mathcal{D}|\Theta^\mathrm{inf}) = & &\prod_{i=1}^{N_\mathrm{relations}} \prod_{j=1}^{N_\mathrm{bins}} \ \ \ \ \ \ \ \   \\
& &\sum_{k=1}^{N_\mathrm{trials}} N_\mathrm{trials}^{-1}  \sum_{l=1}^{N_\mathrm{datasets}} \big( N_\mathrm{datasets}^{-1}  \ \ \ \ \ \ \ \ \ \ \ \ \nonumber \\
 &&  ESN( \tilde{\psi}_{ijk} | \mu(\mathbf{q}_{ijkl}), f_\sigma\sigma(\mathbf{q}_{ijkl}), \epsilon(\mathbf{q}_{ijkl}) ) \big) \nonumber
\end{eqnarray}
Here $\mathbf{q}_{ijkl}$ is shorthand for $q_{16}$, $q_{50}$, and $q_{84}$, the quantiles of the data at a particular mass, which determine the values of $\mu$, $\sigma$, and $\epsilon$ of each epsilon-skew-normal distribution as described above. $ \tilde{\psi}_{ijk} $ is the adjusted set of predicted values from the emulator for a particular relation ($i$) and $z=0$ halo mass ($j$ and $k$ ) given the values of $\Theta^\mathrm{inf}$, which is not quite the same as $\Theta$ as discussed at the beginning of the section.

The data to which we are comparing extend over a large range of masses and presumably different accretion histories, so for each point $\Theta^\mathrm{inf}$ in the parameter space for which we wish to evaluate the likelihood, we employ a grid of $N_\mathrm{trial}\cdot N_\mathrm{bins}=200$ galaxies with different accretion histories covering the range of $M_{h,0}$ from $10^{11} M_\odot$ to $10^{12} M_\odot$, but all with the same value of $\Theta^\mathrm{inf}$. These 200 samples are divided evenly into $N_\mathrm{bins} = 10$ bins in $M_{h,0}$. Within these bins the likelihood of each model is averaged (essentially averaging over different possible accretion histories), then each bin's average likelihood is treated as an independent data point, and these likelihoods are multiplied together. In principle $N_\mathrm{trials}$ could be arbitrarily large, and is only limited by computational power. $N_\mathrm{bins}$ should, however, remain reasonably small and should be set by the typical systematic error in mass. Each time the likelihood is evaluated, the same 200 accretion histories are used. The outer sum over $i$ denotes the $N_\mathrm{relations}=29$ different observational samples at a variety of redshifts to which the model is compared. The inner average over $N_\mathrm{datasets}$ applies when multiple datasets have overlapping coverage. The predicted value of the dependent variable, $y_{\mathrm{pred},ij}$ is compared to the data evaluated at $x_{\mathrm{pred},ijk}$. Note also that this comparison is done in $\log_{10}$-space. 

For datasets in Table \ref{tab:obs} with an explicitly-assumed scatter, $q_{84}$ and $q_{16}$ are taken to be the median $\pm$ the stated scatter rather than the true quantiles of the data. By default the distribution we adopt in the likelihood function, i.e. the error assigned to an observation, is taken to be the population distribution, rather than e.g. the standard error on the median. Implicit in this assumption is that the scatter observed in the population is comparable to, and perhaps even set by, the systematic uncertainty in the variable in question. This assumption is borne out in cases where we plot multiple observational reports of a given scaling relation (see Figures \ref{fig:smhm}, \ref{fig:ssfr}, and \ref{fig:mre}). 

\section{Results} 
\label{sec:results}

With a prior, a likelihood, and a fast way to evaluate it, we can now embark on a the standard procedure to draw samples from the joint posterior distribution $p(\Theta^\mathrm{inf} | \mathcal{D}) \propto p(\Theta^\mathrm{inf}) \mathcal{L} (\mathcal{D}|\Theta^\mathrm{inf})$, thereby fitting the model at up to 4 different redshifts to 8 different galaxy scaling relations: the stellar mass-halo mass relation, the star-forming main sequence, the (gas-phase) mass-metallicity relation, the stellar mass-molecular gas mass relation, the mass-size relation, the Tully Fisher relation, and the relationship between star formation rate and gas velocity dispersion. We use the popular affine-invariant ensemble MCMC sampler emcee \citep{foreman-mackey_emcee_2013} in its parallel tempering mode, which aids the MCMC in converging if there are multiple posterior maxima, where several sets of walkers are run in parallel with different temperatures, i.e. sampling from posterior-like distirbutions where the likelihood is downweighted as the temperature increases $\mathcal{L}(\mathcal{D}|\Theta^\mathrm{inf})^{1/T} p(\Theta^\mathrm{inf})$. We use 13 temperatures, each one spaced a factor of $\sqrt{2}$ higher that the previous one, each with 400 walkers, run for about 50000 iterations. 

\begin{table}
\caption{Largest correlations in the posterior distribution.}
\label{tab:corr}
\centering
\begin{tabular}{lll}
Variable 1 & Variable 2 & Correlation\\
\hline
\hline
$\delta v_\phi$  &  $\alpha_\mathrm{con}$  &  -0.690925775145 \\
$\epsilon_\mathrm{ceil}$  &  $\log_{10}$ $\alpha_r$  &  0.517586893214 \\
$\alpha_{M_h}$  &  $\log_{10}$ $\mu_0$  &  0.5114833758 \\
$\delta \sigma$  &  $\log_{10}$ $\epsilon_\mathrm{ff}$  &  -0.5062732192 \\
$\delta \sigma$  &  $\log_{10}$ $\chi_\mathrm{inj}$  &  -0.498335422386 \\
$\delta r_*$  &  $\log_{10}$ $\epsilon_\mathrm{ff}$  &  -0.494524799007 \\
$\chi_{dlogZ/dlogM}$  &  $\xi_\mathrm{acc}$  &  -0.444850693792 \\
$\delta \sigma$  &  $\xi_\mathrm{acc}$  &  -0.417609635046 \\
$\delta \sigma$  &  $\log_{10}$ $\eta$  &  0.396466878406 \\
$\xi_\mathrm{acc}$  &  $\alpha_{f_g}$  &  0.394964186987 \\
$\alpha_{M_h}$  &  $\alpha_\Sigma$  &  0.383804694949 \\
$\delta Z_g$  &  $\delta M$  &  0.382155536477 \\
$\log_{10}$ $M_Q$  &  $\log_{10}$ $k_Z$  &  -0.380137238756 \\
$\delta M$  &  $\alpha_{M_h}$  &  -0.373905582251 \\
$\delta r_*$  &  $\xi_\mathrm{acc}$  &  -0.367786448618 \\
$\delta \sigma$  &  $\chi_{dlogZ/dlogM}$  &  0.351559006854 \\
$\delta r_*$  &  $\log_{10}$ $\chi_{Z_\mathrm{IGM}}$  &  -0.351400881742 \\
$\log_{10}$ $\eta$  &  $\alpha_{f_g}$  &  -0.343319841545 \\
$\log_{10}$ $\epsilon_\mathrm{quench}$  &  $\alpha_{f_g}$  &  -0.337655782018 \\
$\log_{10}$ $M_Q$  &  $\log_{10}$ $\epsilon_\mathrm{quench}$  &  0.32411469497 \\
$\log_{10}$ $\chi_\mathrm{inj}$  &  $\log_{10}$ $\epsilon_\mathrm{quench}$  &  -0.321362299208 \\
$\log_{10}$ $M_Q$  &  $\log_{10}$ $\alpha_\mathrm{MRI}$  &  -0.319873706753 \\
$\chi_{dlogZ/dlogM}$  &  $\log_{10}$ $\epsilon_\mathrm{ff}$  &  -0.319490839609 \\
$\log_{10}$ $\chi_{Z_\mathrm{IGM}}$  &  $\log_{10}$ $\alpha_{r,*,0}$  &  0.319330567119 \\
$\log_{10}$ $Q_f$  &  $\log_{10}$ $\chi_{Z_\mathrm{IGM}}$  &  0.318563749147 \\
$\delta$ SFR  &  $\delta M$  &  0.317409994256 \\
$\log_{10}$ $M_Q$  &  $\alpha_{f_g}$  &  -0.308359158663 \\
$\xi_\mathrm{acc}$  &  $\alpha_{M_h}$  &  -0.307739361047 \\
$\log_{10}$ $\eta$  &  $\xi_\mathrm{acc}$  &  -0.306755458255 \\
$\chi_{dlogZ/dlogM}$  &  $\log_{10}$ $M_Q$  &  0.303294525319 \\
$\log_{10}$ $Q_f$  &  $\log_{10}$ $\eta$  &  0.302970166563 \\
$\log_{10}$ $\epsilon_\mathrm{ff}$  &  $\log_{10}$ $\chi_{Z_\mathrm{IGM}}$  &  0.301910425097 \\
$\epsilon_\mathrm{ceil}$  &  $\log_{10}$ $Q_f$  &  0.300314976296 \\
$\delta \sigma$  &  $\log_{10}$ $M_Q$  &  0.294728010292 \\
$\log_{10}$ $\epsilon_\mathrm{ff}$  &  $\log_{10}$ $Q_f$  &  0.293989114344 \\
$\delta Z_g$  &  $\log_{10}$ $\mu_0$  &  0.291845453057 \\
$\chi_{dlogZ/dlogM}$  &  $\log_{10}$ $\alpha_\mathrm{MRI}$  &  -0.289593339158 \\
$\epsilon_\mathrm{ceil}$  &  $\log_{10}$ $\eta$  &  0.286666051006 \\
$\log_{10}$ $\chi_\mathrm{inj}$  &  $\log_{10}$ $\epsilon_\mathrm{ff}$  &  0.286587872447 \\
$\log_{10}$ $\epsilon_\mathrm{ff}$  &  $\log_{10}$ $\alpha_\mathrm{MRI}$  &  0.284022258999 \\
\hline

\end{tabular}\par
\bigskip
\end{table}

\begin{figure*}
\centering
\includegraphics[width=6.5in]{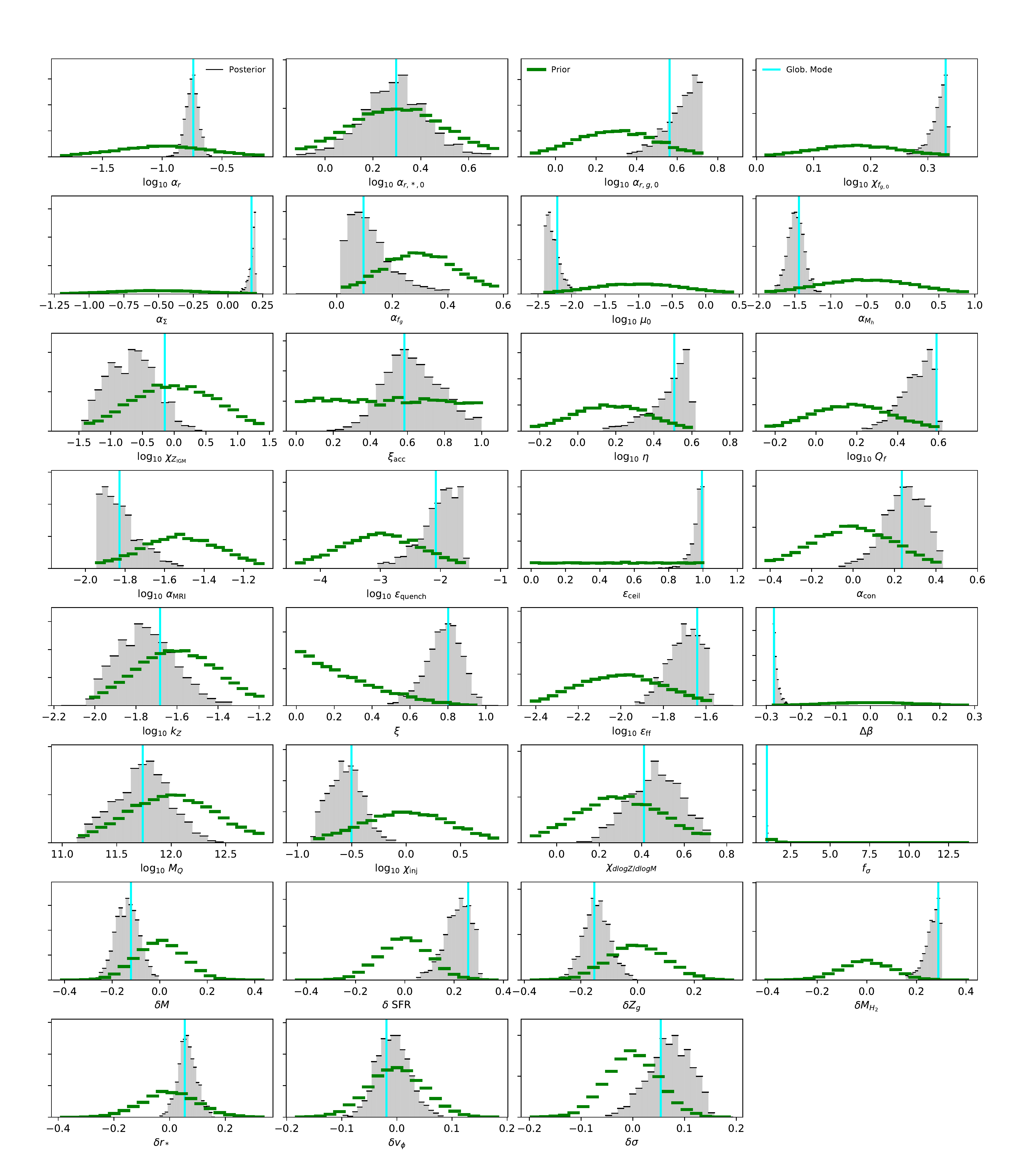}
\caption{Marginal Distributions of the Posterior. Each panel shows the prior (green), the posterior (gray), and an estimate of the global posterior mode (vertical cyan lines)}
\label{fig:marginals}
\end{figure*}

\begin{figure*}
\centering
\includegraphics[width=7in]{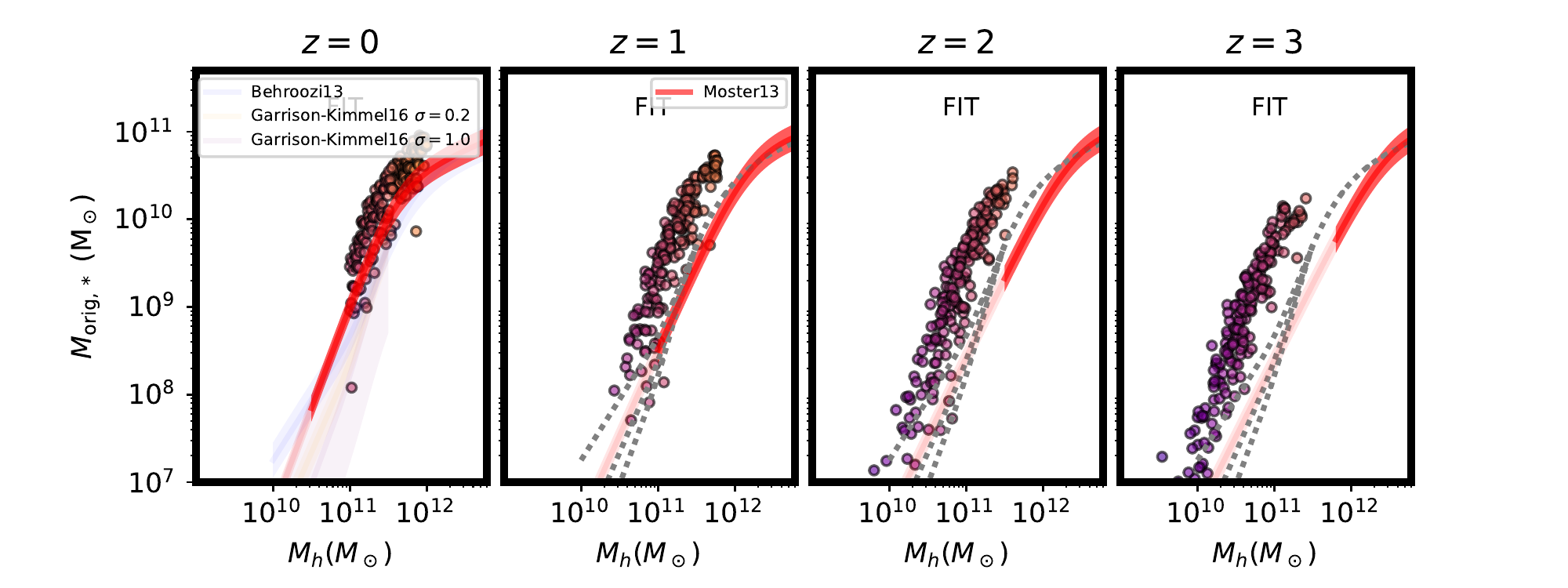}
\caption{The stellar mass-halo mass relation at a variety of redshifts. Each point is a simulated galaxy colored by its $z=0$ halo mass, while the shaded regions represent the mean relation and scatters from \citet{moster_galactic_2013} (red). Also shown are versions of this relation from \citet{garrison-kimmel_organized_2017} and \citet{behroozi_average_2013}.}
\label{fig:smhm}
\end{figure*}

\begin{figure*}
\centering
\includegraphics[width=7in]{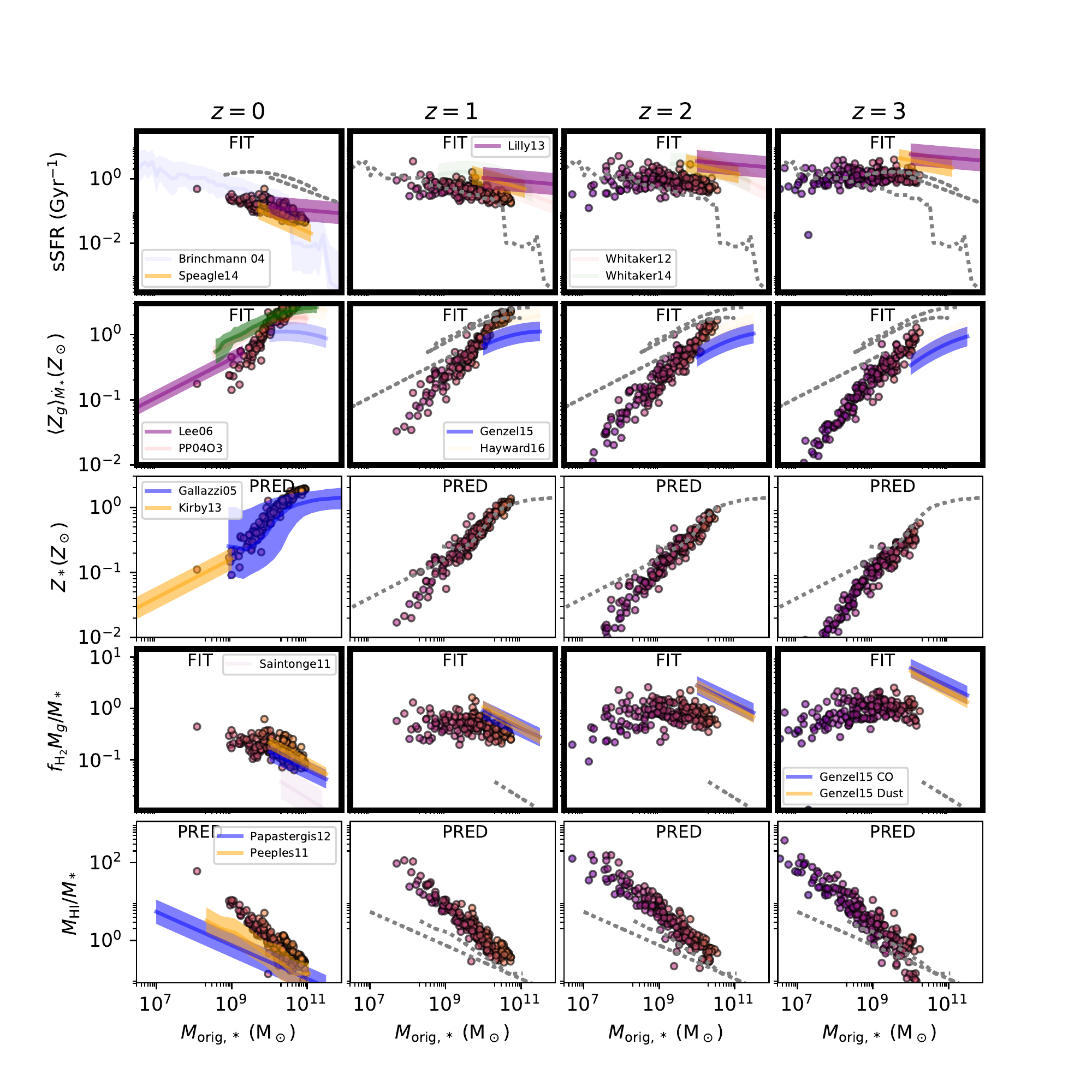}
\caption{Comparison with gas-related galaxy scaling relations. As in Figure \ref{fig:smhm}, model galaxies are shown as points colored by their $z=0$ halo mass. Each column shows a different redshift, and each row shows a different quantity plotted against the galaxy's stellar mass. Data from \citet{brinchmann_physical_2004}, \citet{speagle_highly_2014-1}, \citet{whitaker_star_2012}, \citet{whitaker_constraining_2014}, \citet{lee_extending_2006-2}, \citet{genzel_combined_2015}, \citet{hayward_how_2017}, \citet{kewley_metallicity_2008}, \citet{tremonti_origin_2004}, \citet{gallazzi_ages_2005}, \citet{kirby_universal_2013}, \citet{saintonge_cold_2011}, \citet{papastergis_direct_2012}, and \citet{peeples_constraints_2011-1} are shown for comparison.}
\label{fig:ssfr}
\end{figure*}

\begin{figure*}
\centering
\includegraphics[width=7in]{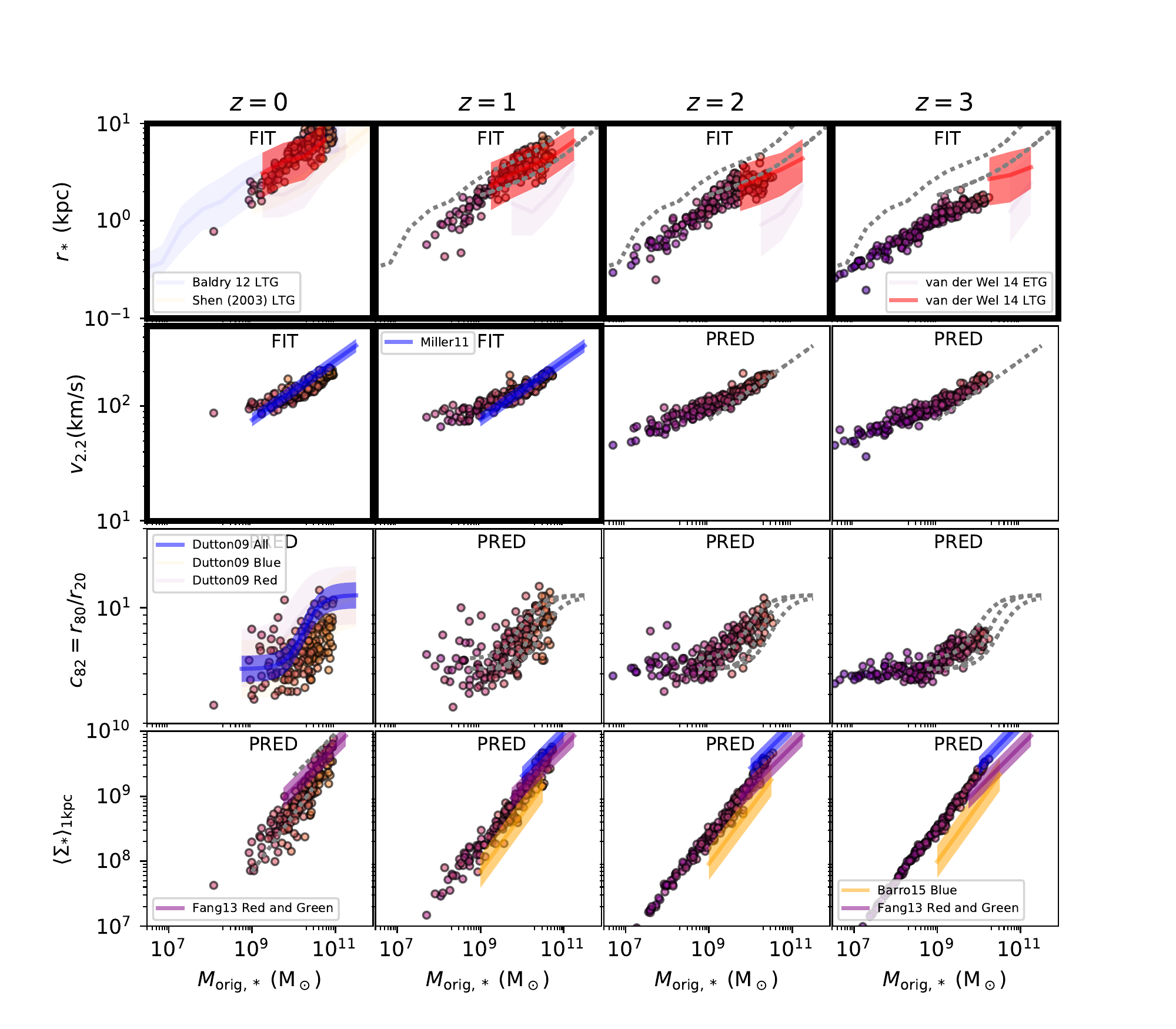}
\caption{A comparison with galaxy scaling relations related to the radial distribution of stars. Model galaxies are shown as points colored by their $z=0$ halo mass, observational relations are shown as colored bands. Data from \citet{baldry_galaxy_2012}, \citet{shen_size_2003}, \citet{van_der_wel_3d-hst+candels_2014}, \citet{miller_assembly_2011}, \citet{dutton_origin_2009}, \citet{fang_link_2013}, and \citet{barro_structural_2017} are shown for comparison.}
\label{fig:mre}
\end{figure*}

\begin{figure*}
\centering
\includegraphics[width=7in]{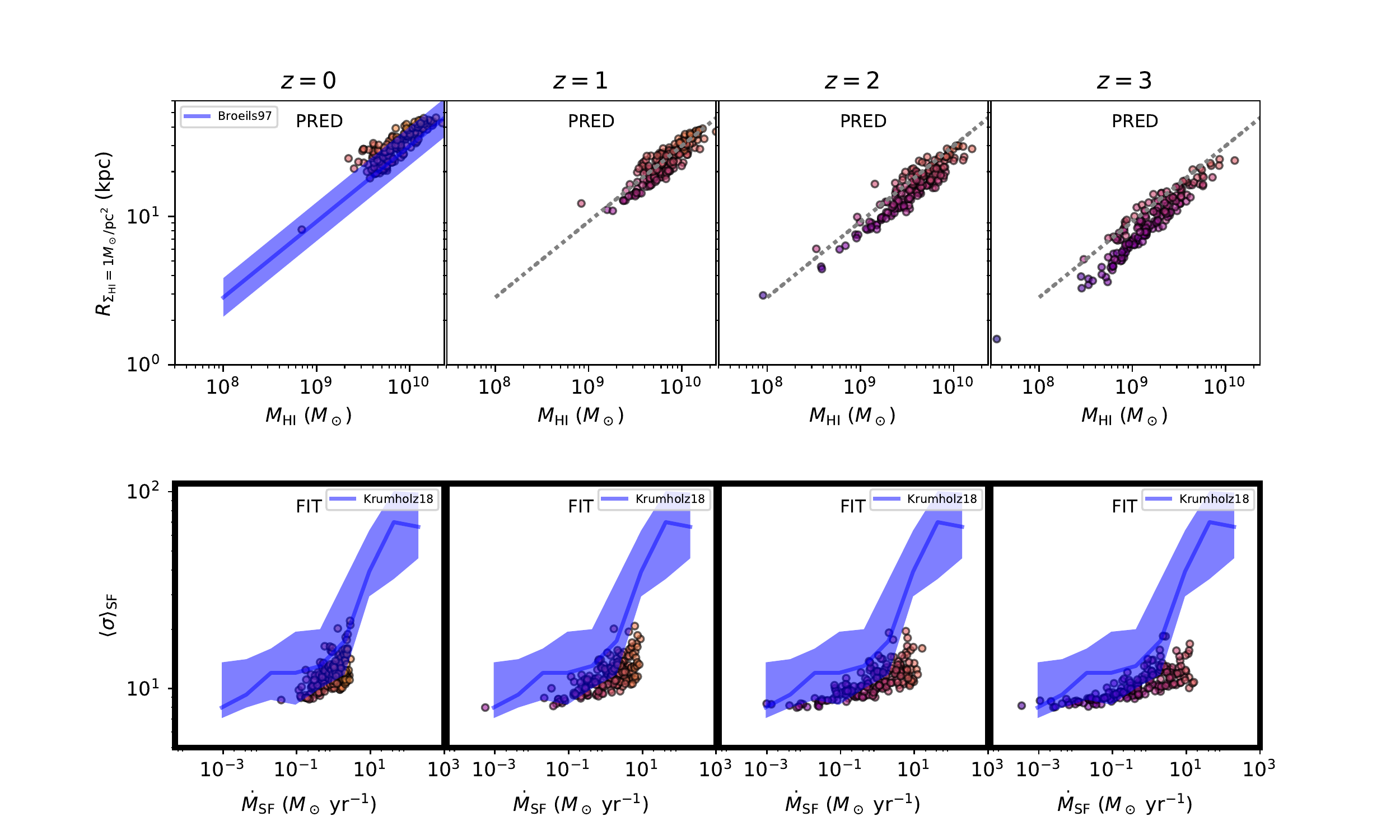}
\caption{A comparison with galaxy scaling relations that do not depend explicitly on stellar mass. Model galaxies are shown as points colored by their $z=0$ halo mass, observational relations are shown as colored bands. Data from \citet{broeils_short_1997} and \citet{krumholz_unified_2018} are shown for comparison.}
\label{fig:nonmst}
\end{figure*}

The marginal posterior distributions of each parameter are shown in Figure \ref{fig:marginals}. The gray histograms are the marginal 1D posterior distributions themselves. For comparison, the green line shows the prior of each parameter, i.e. the narrow prior actually used in the inference process, not the distribution from which the training set was drawn. In most, though not all, cases the prior and the posterior are substantially different, indicating that the set of observations to which we are comparing provide meaningful constraints on the parameters in question. Also shown are estimates for the posterior mode of each run, plotted as light blue vertical lines. These modes are estimated as the posterior sample whose 30th-nearest-neighbor is least distant. Distances are measured in the same space used by the MCMC, namely one in which the logarithm of many of the parameters has been taken, and the parameter values have been linearly scaled by their interquartile range.

We see first that it is quite normal for the posterior distribution to be concentrated right at the hard boundary imposed by the $2\sigma$ cutoff in the prior. The solution favored by this combination of physical model, priors, and data has a number of interesting features. The feedback parameters are such that massive galaxies have very low mass loading factors (since $\mu_0 \ll 1$), but low-mass galaxies need strong feedback, given the large negative values of $\alpha_{M_h}$. These winds are highly metal-enhanced (as seen by the large values of $\xi$). The reaccretion of metals from winds is also quite important, given the large inferred value of $\xi_\mathrm{acc}.$ The galaxies also begin life with more stellar mass and gas mass than one might expect, given the large negative values of $\Delta \beta$ and positive values of $\chi_{f_g}$. The accretion distribution is taken to be quite flat in radius (given the large value of $\alpha_r$). Gravitational instability is quite important in this posterior distribution, $Q_f$ is quite large, favoring easy movement through the disk, while a simple $\alpha$ viscosity (represented by the value of $\alpha_\mathrm{MRI}$ is disfavored in importance.

The 1D marginal distributions of the posterior already reveal quite a lot about how the fit works, but we can gain additional insight by looking at the correlations between each physical variable. The largest 40 correlations in absolute value, corresponding to the top $\sim$10\% of correlations among all possible pairs of parameters, are shown in Table \ref{tab:corr}. The variables allowing systematic offsets in the observable quantities are well-represented among these correlations, giving us a quick sense of which observables are most strongly affected by which physical parameters (e.g. $v_\phi$ and $\alpha_\mathrm{con}$, or $\sigma$ and $\chi_{inj}$). There are also important tradeoffs between the physical parameters, chief among them $\epsilon_\mathrm{ceil}$ and $\alpha_r$. Essentially the larger $\alpha_r$ is, the more of the accretion profile ends up falling at very large radii, in which case a larger value of $\epsilon_\mathrm{ceil}$ is necessary to supply the gas needed to form the observed stars.

Figures \ref{fig:smhm}, \ref{fig:ssfr}, \ref{fig:mre}, and \ref{fig:nonmst} show the results of re-simulating samples from the posterior distribution. In particular, we draw 180 samples\footnote{Note that not all of these models successfully run to $z=0$.} from the posterior distribution. These plots are organized as follows. Each row corresponds to a different observational relation, and each column shows a different redshift. The points show the re-simulated sample of galaxies, while the colored bands show the observational data. In cases where observations are not available at a given redshift, and where the galaxies extend outside the mass range of the observations, these points are genuine predictions. Entire panels where the models are not being compared directly to observations are labelled as predictions, `PRED,' while panels where at least some models are being compared to data are labelled `FIT' and emphasized with a thicker black border. The general impression left by these figures is that the parameters identified as part of the posterior distribution, despite all the estimates that go in to generating it, does a reasonable job of reproducing the data to which it was fit. The points in each figure are colored by their $z=0$ halo mass $M_{h,0}$.

\begin{figure*}
\centering
\includegraphics[width=7in]{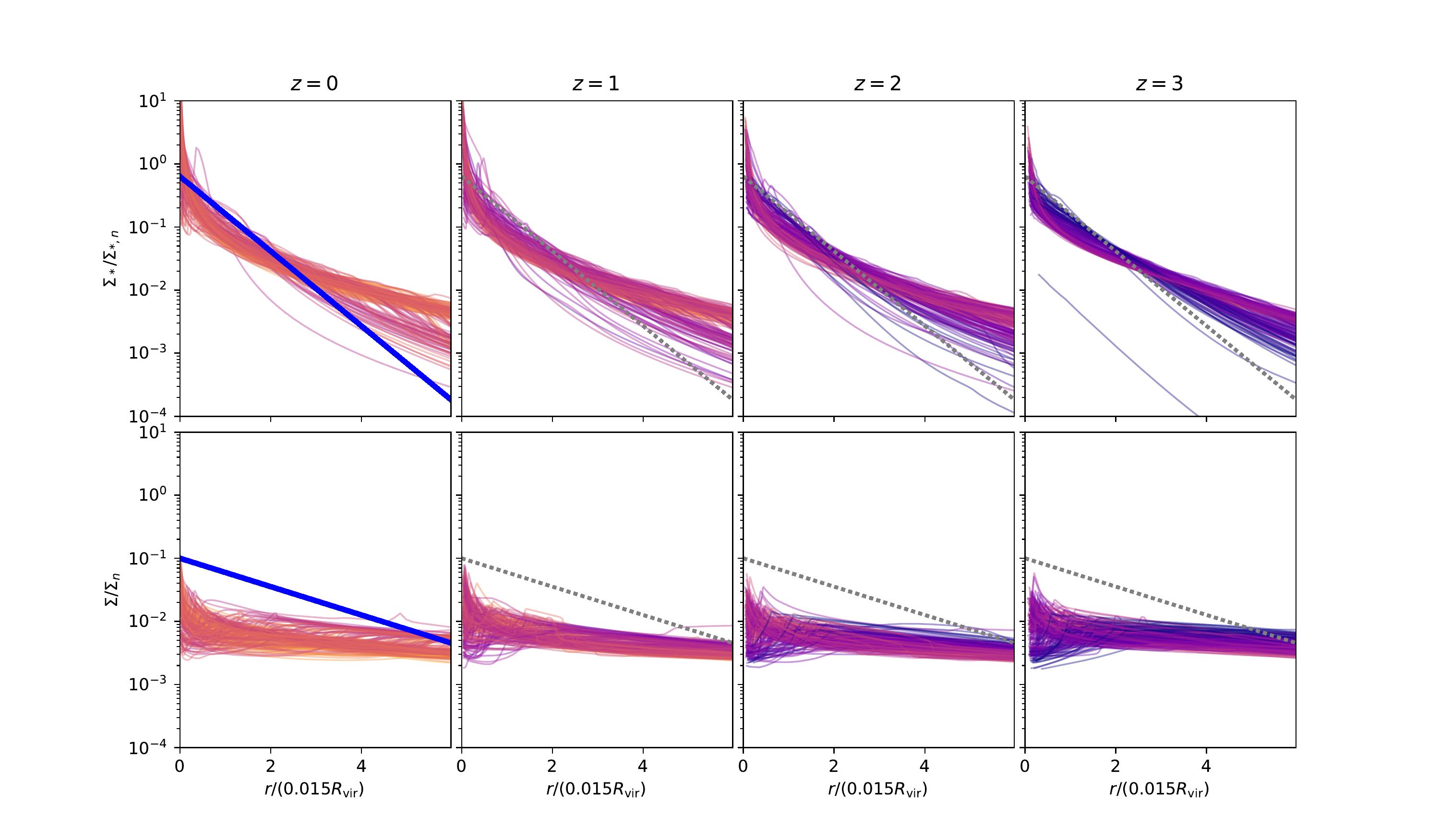}
\caption{Scaled radial profiles of gas and stars. Model gas and stellar surface densities are scaled following \citet{kravtsov_size-virial_2013} (see text) and plotted as colored lines, with purple representing lower values of the halo mass at that epoch, and yellow representing higher halo masses. The blue line is the $z=0$ relation suggested by \citet{kravtsov_size-virial_2013} - the dotted line is just this line repeated at higher redshifts.}
\label{fig:kravtsovStruct}
\end{figure*}

In fitting the stellar mass-halo mass relation (Figure \ref{fig:smhm}), we see that the simulations fit the \citet{moster_galactic_2013} relation reasonably well at $z=0$, but across all redshifts the stellar masses moderately exceed the \citet{moster_galactic_2013} values. This is both concerning and interesting. Because we rely on the stellar mass halo mass relation to ``regularize'' the fit, i.e. to prevent the model from having too much freedom to adjust the x-axis values of all the observed relations, we include the \citet{moster_galactic_2013} constraints at all redshifts at all masses. Of course, this is an extrapolation, particularly at high redshift, where that model has not been constrained with low-mass galaxies. Hence we are seeing that the MCMC is willing to sacrifice a good fit to \citet{moster_galactic_2013} (by using $\Delta \beta < 0$) in favor of better-fitting all of the other data. In the end this is compelling evidence that \citet{moster_galactic_2013} underpredicts the stellar mass of low-mass halos at high redshift. This is in line with other lines of evidence from dynamical mass measurements \citep{burkert_angular_2016} and simplified modeling at higher redshifts \citep{tacchella_redshift-independent_2018}. 
Figure \ref{fig:ssfr} shows relations associated with the global gaseous processes, i.e. star formation rate, metallicity, and gas fractions. Despite the general reasonableness of the fit, there are several areas of tension, especially the mass-metallicity relation. This is likely in part due to the uncertainty in the $z=0$ relation, where different metallicity callibrations lead to different slopes and normalizations \citep{kewley_metallicity_2008}. The likelihood function treats each relation as plausible, so a mixture model may not be sufficient to account for the systematic uncertainties in this relation. Despite this tension, there is also the remarkable success that the $Z_*-M_*$ relation, despite not being included in the likelihood, fits almost perfectly. 
Also notable in Figure \ref{fig:ssfr} is a common feature of the fitting procedure. At high redshifts, the observations are restricted to high-mass objects, whereas our models are limited to galaxies that are unlikely to live in groups or clusters at $z=0$ in order to minimize the effects of mergers and quenching. The observations therefore provide little constraint on the models at the highest redshifts. Either improved prescriptions for mergers and quenching, or some principled means to use higher-mass models at $z\sim3$ while ignoring the results of those models at lower redshift will be required to fully leverage the high-redshift data.

Figure \ref{fig:mre} shows relations associated with the stellar structure of the galaxy, namely the half-mass radius, the Tully-Fisher relation, the concentration, and the central stellar density. Here the fits are remarkably good, with the possible exception of the low $\Sigma_1$ part of the $\Sigma_1-M_*$ relation at high $z$. In part this may be attributed to the disconnect between the \citet{barro_structural_2017} and \citet{fang_link_2013} relations. Regardless, this is a remarkable success of the model because neither $\langle\Sigma_*\rangle_\mathrm{1\ kpc}$ nor $c_{82}$ were used in the fit.

\begin{figure*}
\centering
\includegraphics[width=7in]{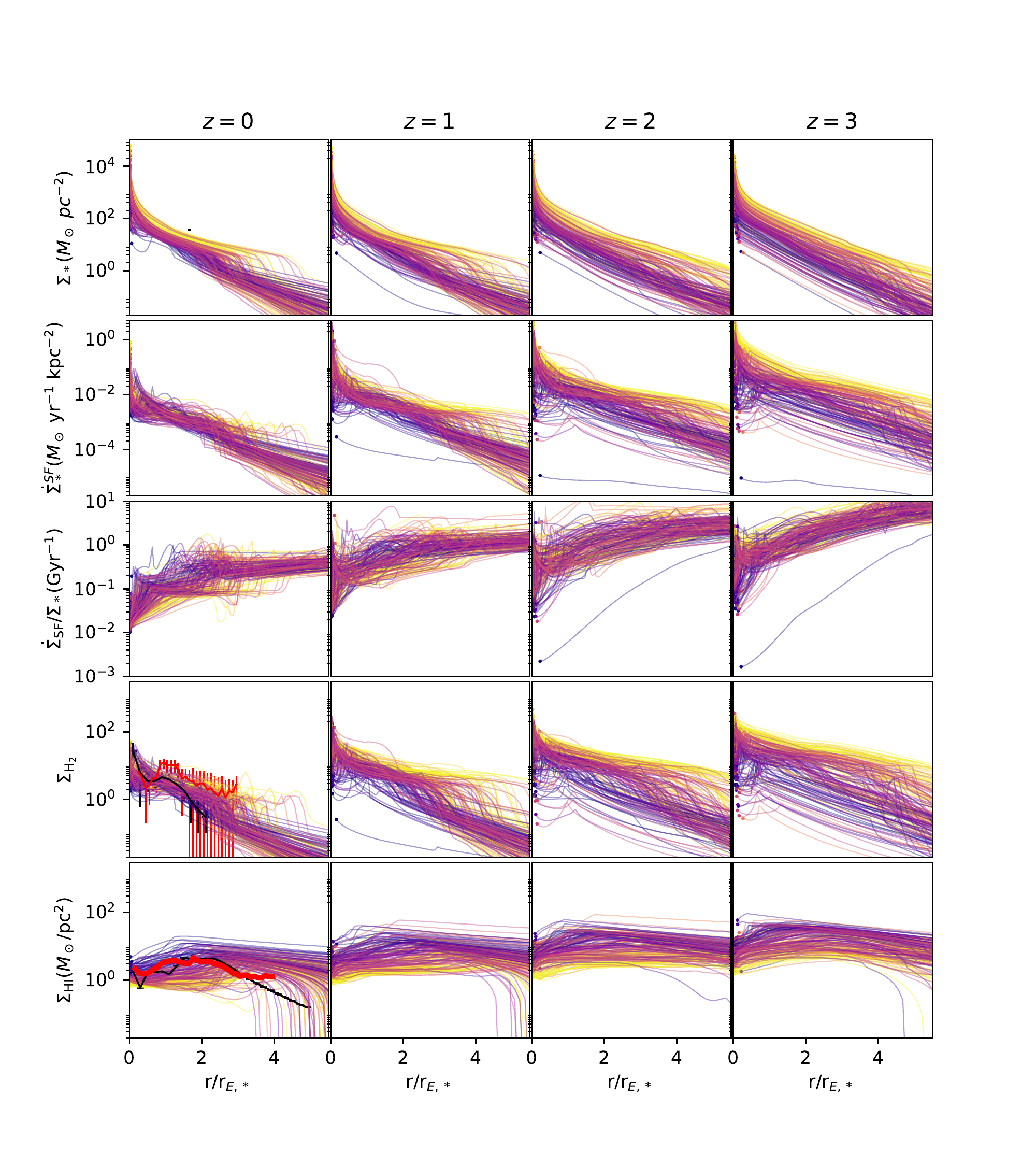}
\caption{Radial distributions. Various quantities for models drawn from the posterior distribution are shown as a function of $r_{E,*}$, defined as the radius containing half of a galaxy's stellar mass. A simple comparison with Milky Way data from \citet{nakanishi_three-dimensional_2003} and \citet{nakanishi_three-dimensional_2006} for HI and H$_2$ respectively is shown in black, while the red lines in those plots are derived from Herschel data in \citet{pineda_herschel_2013}. }
\label{fig:cols1}
\end{figure*}

\begin{figure*}
\centering
\includegraphics[width=7in]{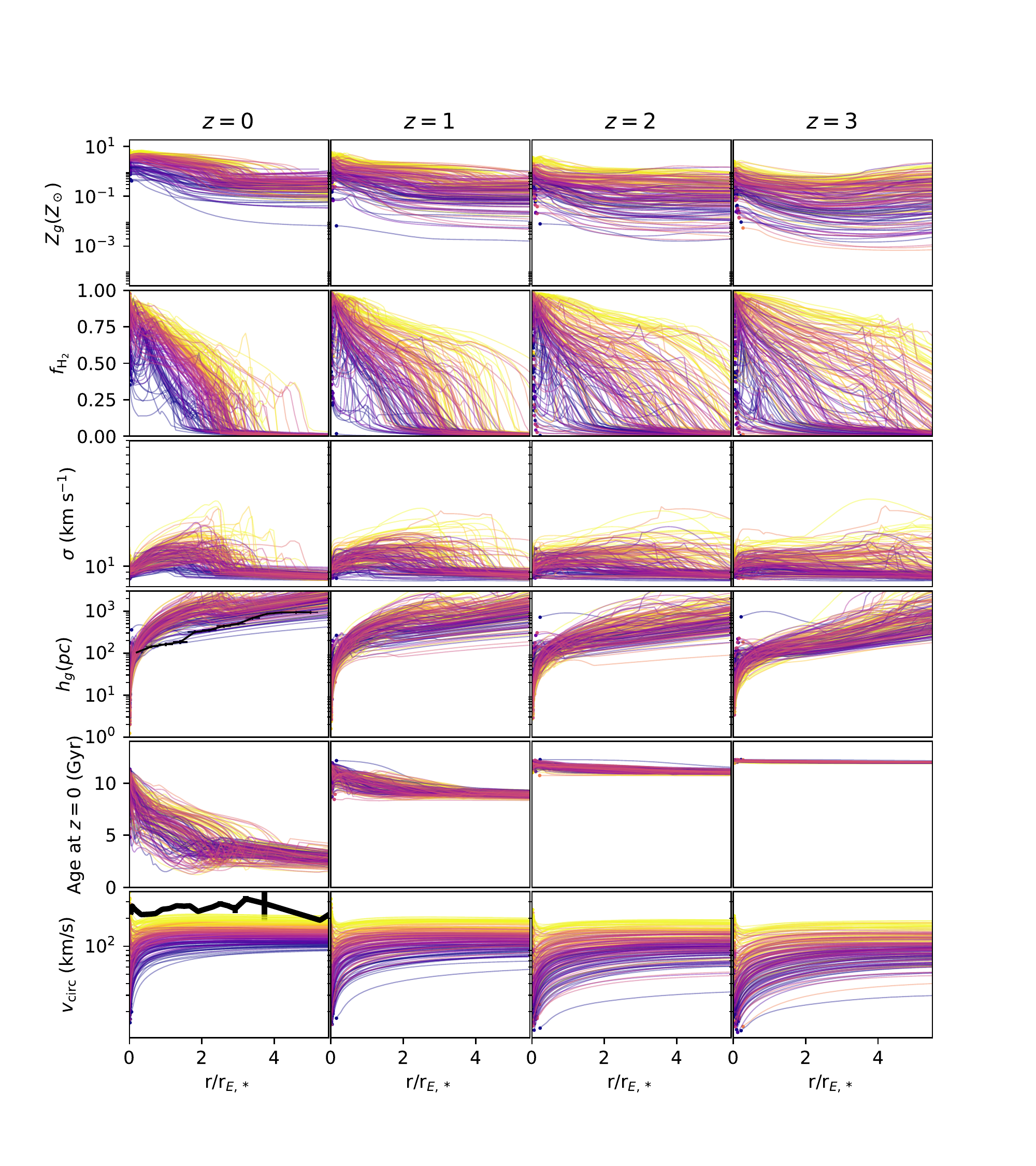}
\caption{More radial distributions. The scale height data for the Milky Way are from \citet{nakanishi_three-dimensional_2003}, and the rotation curve data are from \citet{bhattacharjee_rotation_2014}}
\label{fig:cols2}
\end{figure*}

The values fit for the Tully-Fisher relation agree reasonably well with the data as reported by \citet{miller_assembly_2011}. Given that the model has some slight tension with the stellar mass-halo mass relations inferred by abundance matching (see Figure \ref{fig:smhm}), it is perhaps surprising that the models also agree with the circular velocity measured at sufficiently large radii, but as we saw at the top of Table \ref{tab:corr}, the model has some freedom to adjust its concentration as a tradeoff to this velocity, and it does so by increasing the concentration, consistent with expectations for some baryonic contraction in these modestly massive galaxies.

Rounding out the plots of quantities used in the fitting procedure is Figure \ref{fig:nonmst}, which shows relations that do not depend directly on stellar mass, in particular the strong correlation observed by \citet{broeils_short_1997} between a galaxy's mass in HI and the radius at which the HI surface density first drops to $1 M_\odot/\mathrm{pc}^2$. Once again we find that the agreement between the data and the model is quite good despite not being included in the fit. 

 Also shown is a comparison to the compilation of data from \citet{krumholz_unified_2018}, in particular the correlation between star formation rate and gas-phase velocity dispersion. Although this relation has been claimed as evidence for the importance of stellar feedback in driving velocity dispersion, \citet{krumholz_is_2016} and \citet{krumholz_unified_2018} have shown that a similar correlation arises from the viscous transport associated with gravitational instability, and that a hybrid model including both effects does the best job of explaining the data. The full evolutionary calculations we consider here have at their core the same physics, but include more details. The models presented here do not quite reach the large observed values of $\sigma$ seen in the data, which is probably among the largest tensions between our model and the data. These models are perfectly capable of producing the necessary large velocity dispersions, by including gravitational instability which kicks in at a large value of $Q$. Indeed the best fit models do include large values of $Q_f$, but also large values of $\alpha_r$, which sends much of the gas to larger radii and moderates the resulting gravity-driven turbulence. Essentially, in order to fit the rest of the data, the model is willing to live with slightly lower values of $\langle \sigma \rangle_\mathrm{SF}$ in this diagram.

We now turn to the issue of validating the model, that is, comparing it to relevant data not used in the fits. We have already seen several instances of this in Figures \ref{fig:ssfr} and \ref{fig:mre}, namely the relations between $c_{82}$ and $M_*$, $M_{\mathrm{HI}}/M_*$ and $M_*$, $Z_*$ and $M_*$ and $\langle\Sigma\rangle_\mathrm{1 kpc}$ and $M_*$. According to \citet{dutton_origin_2009}, the relationship between mass and concentration essentially follows a gradual transition from profiles consistent with exponential discs to profiles consistent with de Vocaleurs profiles \citep{de_vaucouleurs_outsiders_1978} as the stellar mass crosses the characteristic quenching mass. Interestingly, both blue and red galaxies follow similar relations, so the relationship between the quenching of star formation and the morphological bimodality of galaxies is not one-to-one, though both transformations are associated with the same mass scale. Recall that the emulator did not perform well when predicting values of $c_{82}$, so it was removed from the likelihood function and not used in the fitting process. Nonetheless the models drawn from the posterior distributions do a decent job of reproducing the features of the data, namely a transition from low-concentration disc-like profiles to higher-concentration, higher-Sersic-index profiles at stellar masses around $10^{10} M_\odot$.

Rather than comparing simple statistics extracted from the radial profiles of the stars, we can explicitly compare the stellar profiles produced in our simulations to observational data. \citet{kravtsov_size-virial_2013} provides a convenient means to do so by pointing out that galaxies across 6 decades in stellar mass appear to follow a consistent relationship between an appropriately-normalized stellar column density and a radius scaled to the estimated Virial radius of their dark matter halo. Defining $r_n=0.015 R_\mathrm{Vir}$ and $\Sigma_n=0.448 M/r_n^2$, $M$ being the total mass of gas or stars in the profile, the stellar profiles fall along a relation reasonably approximated as exponential between an $r/r_n$ of about 1 and 3. This exponential profile is shown as a black line, along with model profiles in Figure \ref{fig:kravtsovStruct} colored according to their instantaneous halo mass, just as in earlier figures, with yellow indicating high masses, and purple low. Not only are the models in reasonable agreement with the suggested scaled exponential stellar profile at $z=0$, at least within a few scale lengths, comparing Figure \ref{fig:kravtsovStruct} with Figure 2 of \citet{kravtsov_size-virial_2013} shows that the deviations of real galaxies from this exponential profile are also consistent with the model: at both $r/r_n<1$ and $r/r_n>3$, the model profiles and the data exceed the exponential profile. The data do not quite reach $\Sigma_*/\Sigma_n \sim 1$ as our models do, but this is precisely where the dynamics of our model are not reliable given that $\sigma_*/v_\mathrm{circ}$ is no longer far below unity.

The lower panel compares the surface density of gas to the exponential profile suggested by \citet{kravtsov_size-virial_2013}, which he explains may be rescaled directly from the universal exponential profile of gas in disc galaxies proposed by \citet{bigiel_universal_2012}. Here the model galaxies fall substantially below the observed relation as a result of the large scale length of the cosmologically accreting gas. This suggests that perhaps improved fits could be made by allowing more freedom in the radial profile of cosmological accretion.  

Figures \ref{fig:cols1} and \ref{fig:cols2} show the raw profiles from which other quantities shown thus far have been derived. For the sake of a visual comparison, we have also included a few datasets from the Milky Way for comparison with the massive galaxies (which appear as yellow lines). The HI and H$_2$ profiles shown in black lines are taken from \citet{nakanishi_three-dimensional_2003} and \citet{nakanishi_three-dimensional_2006} respectively. The former also compiled the estimates for the gas scale height as a function of radius shown in Figure \ref{fig:cols2}. The red lines for the HI and H$_2$ profiles are derived from Herschel data as presented in \citet{pineda_herschel_2013}. The rotation curve of the Milky Way shown in Figure \ref{fig:cols2} was compiled by \citet{bhattacharjee_rotation_2014}. Once again the models look reasonable in comparison to these data, despite not being tuned to fit these particular datasets.

These plots also offer some further insights into the behavior of the model galaxies. Based on the evolution of the specific star formation rate $\dot{\Sigma}_\mathrm{SF}/\Sigma_*$ (Figure \ref{fig:cols1}) and the average $z=0$ age of the stellar population (Figure \ref{fig:cols2}), we see that galaxies slow down their star formation from the inside out. This is consistent with the measurements presented by \citet{tacchella_evidence_2015} for massive star-forming galaxies at $z\sim2$ at a mass range somewhat larger than the models under consideration here, and \citet{nelson_where_2016} at lower masses and redshifts. The cause in the models of these profiles of $\dot{\Sigma}_\mathrm{SF}/\Sigma_*$ and stellar age is not completely clear. There are likely several factors that contribute: first, as discussed in \citet{forbes_balance_2014}, the shutoff of radial transport of gas by gravitational instability once the column density falls below what is necessary to sustain $Q\sim 1$ tends to quench galaxies from their centers. This effect is mitigated slightly by a radial profile of gas accretion that extends to the center of the galaxy, as well as the non-negligible radial transport from other sources parameterized as $\alpha_\mathrm{MRI}$. Another effect which may contribute is the transport of stars themselves via spiral arms. On average this effect pushes stars to smaller galactic radii regardless of what the gas is doing, leading to larger values of the denominator when calculating sSFR, namely $\Sigma_*$, with only secondary effects on the numerator.

Metallicity gradients are generally quite flat in star-forming disc galaxies out to at least $\sim 2$ optical radii \citep{werk_metal_2011}. The models reproduce this behavior in high-mass galaxies, but moving to masses slightly below the Milky Way's or to higher redshifts we see that this is not universally true. Lower-mass galaxies can have jumps in their metallicity profiles corresponding to the radii to which metals produced near the center of the galaxies can be diffused or advected through the disc. The galaxy outskirts are not totally devoid of metals owing to the effects of the galactic fountain, but they do not appear to keep up with the in-disc transport, which after all, is operating in gas with much higher densities. These modest dropoffs in metallicity beyond a few effective radii are not inconsistent with the data, but it is difficult to make that measurement so far out.

\section{Summary}
\label{sec:summary}
Despite rapid progress in the hydrodynamical simulation of galaxies over the past few years, much remains fundamentally unknown about the physics of how galaxies operate. At the same time, rich new datasets from IFU surveys have expanded our view of local galaxies from the single central fiber of SDSS. These two facts together point to the usefulness of a flexible, reasonably inexpensive, physical model for the evolution of galaxies that can still make predictions for galaxy properties resolved in radius.

In this work we have presented a first step towards this goal. We have developed a code to evolve discs under the simplifying assumptions of axisymmetry and thinness. The computational elements are annuli spaced logarithmically in radius, with each annulus containing the surface density, velocity dispersion, and $\alpha$ and Fe abundances for gas and stellar populations in uniformly-spaced age bins. In addition, each stellar population has a separate vertical (out-of-plane) and radial (in-plane) velocity dispersion. These quantities are evolved under non-trivial treatments of star formation (regulated by local molecular gas content), galactic winds, cosmological accretion, radial mixing of metals, an accounting of the galaxy's rotation curve, the dynamical heating and radial transport of stars by spiral arms, and the driving of interstellar turbulence and gaseous radial transport by gravitational instability.

Throughout, we take the approach of parameterizing uncertain physical ingredients. These parameters may then be constrained by comparing the model to data with a Bayesian approach. Doing so naively turns out to be prohibitively expensive because the evolution of a single galaxy, while many orders of magnitude cheaper than a cosmological zoom-in simulation, still requires of order 10 minutes depending on the values of the physical parameters. We have therefore developed an emulator that can quickly predict a small set of quantities calculable from the simulations given a point in parameter space. This allows us to run MCMCs in a reasonable amount of time. 

The best fit model employs metal-enhanced galactic winds, and outflows that are modest overall. Cosmologically accreting material is distributed with a large scale length, and gravitational instability plays a substantial role in redistributing gas and stars. Systematic offsets in each observable quantity are explicitly included in the fit, and these parameters are well-represented among the largest 2D correlations between different posterior quantities, which provides an estimate of which observable quantities are most strongly influenced by which parameters in the vicinity of the best-fit.

Overall the model does a good job at reproducing the data to which it was fit. Despite the large number of parameters, this is a non-trivial achievement. The model was required to fit 11 different galaxy scaling relations at up to 4 different redshifts, and in practice many of these observables are primarily influenced by the same handful of parameters. We have also shown reasonable agreement between the model and observations not used at all in the fits.

We suggest that the emulator technique, suitably applied, will be extremely helpful in employing flexible but computationally non-trivial models to fit an increasingly rich range of observational data. We anticipate that the general-purpose models that we have constrained here will be helpful for understanding a wide range of problems, from the origin of scatter in galaxy scaling relations and behaviors of disc galaxies across the star-forming main sequence, to the structure of disc galaxies in terms of their gas, stars, angular momentum and metals, and even cosmological distribution functions, e.g. the probability density of HI column densities originating in galactic discs, or CO intensity mapping.

\section*{Acknowledgements}
JCF is supported by an ITC Fellowship, and would like to thank Joel Primack, Peter Behroozi, Aaron Dutton, Benedikt Diemer, Andrey Kravtsov, Sedona Price, Kevin McKinnon, Hui Li, Rachel Bezanson, Arjen van der Wel, and Sandro Tacchella for helpful conversations, and Lars Hernquist for his support. This work was aided immensely by a number of publicly available resources: NASA ADS, arXiv, numpy, scipy, matplotlib, sklearn, glue, the GNU scientific library, git, and the Bolshoi merger trees. We thank those responsible for maintaining and sharing these resources. Simulations were carried out on the Odyssey cluster supported by the FAS Division of Science, Research Computing Group at Harvard University, and in part on the UCSC supercomputer Hyades supported by NSF grant AST-1229745. MRK acknowledges support from the Australian Research Council's Discovery Projects and Future Fellowship funding schemes (awards DP160100695 and FT180100375).

\bibliography{/Users/jforbes/updatingzotlib}

\appendix

\section{Numerical computation of rotation curve integrals}
\label{app:rotcurve}
Having re-written the equations for $v_{\phi, \mathrm{disc}}^2$ as in equation \ref{eq:vphidisc}, the integrals can be pre-computed at the beginning of each simulation so long as the grid structure does not change over the course of the simulation. In particular, the influence of disc material in the $i$th cell on the circular velocity in the $j$th cell is
\begin{equation}
V_{ij} = \int_0^\infty dk J_1( k r_j) k \int_{r_{i-1/2}}^{r_{i+1/2}} J_0 (k r') r' dr'
\end{equation}
Depending on whether $r_j$ is less than, greater than, or within the $i$th cell, these integrals have closed-form solutions as sums of Elliptic integrals. During the simulation, the contribution to the circular velocity from the disc is simply computed as
\begin{equation}
\label{eq:vphidiscsum}
v_{\phi, \mathrm{disc},j}^2 = \sum_i (\Sigma_i + \Sigma_{*,i}) V_{ij},
\end{equation}
with $V_{ij}$ pre-computed.

In practice, computing $v_{\phi, \mathrm{disc}}^2$ this way and subjecting the disc to transport via gravitational instability, leads to unphysical grid-scale oscillations in both the column density and the rotation curve. Small perturbations in the surface densities leads to corresponding perturbations in the rotation curve. Gravitational instability acts to keep $Q \propto v_\phi/\Sigma$ constant, and so enhances the original perturbations in $\Sigma$. This numerical instability is therefore the result of the simulation attempting to enforce a broad ansatz, i.e. $Q \ga 1$ as precisely true in every annulus. 

To suppress unphysical oscillations in $v_\phi$ and $\Sigma$, we artificially suppress small-scale modes of $\Sigma_i$ and $\Sigma_{i,*}$ when computing equation \ref{eq:vphidiscsum}. In particular, $\Sigma_i$ and $\Sigma_{i,*}$ in that equation are replaced with the inverse Fourier transforms of
\begin{equation}
\label{eq:ifft}
\tilde{\Sigma}(k) = \Sigma_\mathrm{FFT}(k) e^{-(k/k_\mathrm{lim})^{n_\mathrm{lim}}}
\end{equation}
and
\begin{equation}
\label{eq:fft}
\tilde{\Sigma}_*(k) = \Sigma_{*,\mathrm{FFT}}(k) e^{-(k/k_\mathrm{lim})^{n_\mathrm{lim}}}
\end{equation}
where $\Sigma_\mathrm{FFT}$ and $\Sigma_{*,\mathrm{FFT}}$ are the Fast Fourier Transforms of the gas and stellar surface densities. With the appropriate choice of $k_\mathrm{lim}$ and $n_\mathrm{lim}$, high-$k$ (i.e. small-scale) oscillations in the surface densities are suppressed enough that the numerical instability does not develop.

This procedure has the adverse effect of producing locations in the rotation curve with discontinuous first derivatives. These in turn can affect the numerical stability of the equations governing mass transport. To smooth out these discontinuities, we additionally average the full rotation curve (including contributions from all three components) in a fixed windows of 40 cells. This essentially guarantees that regardless of the accretion history or size scale of the galaxy being simulated, the simulation will run smoothly at least as far as the rotation curve is concerned.

\section{Emulator Validation}
\label{app:modelcompare}
In order to assess how well the models do at reproducing the data, we employ a combination of visual and quantitative metrics. Visually we can examine the residuals and any trends they may have with particular variables. For the most part, the residuals are centered about zero and there are no visible trends with $M_{h,0}$. A few variables have larger scatters about zero at the low-mass end of the fits. 

\begin{figure}
\centering
\includegraphics[width=4in, trim={1cm 0 0 0}, clip]{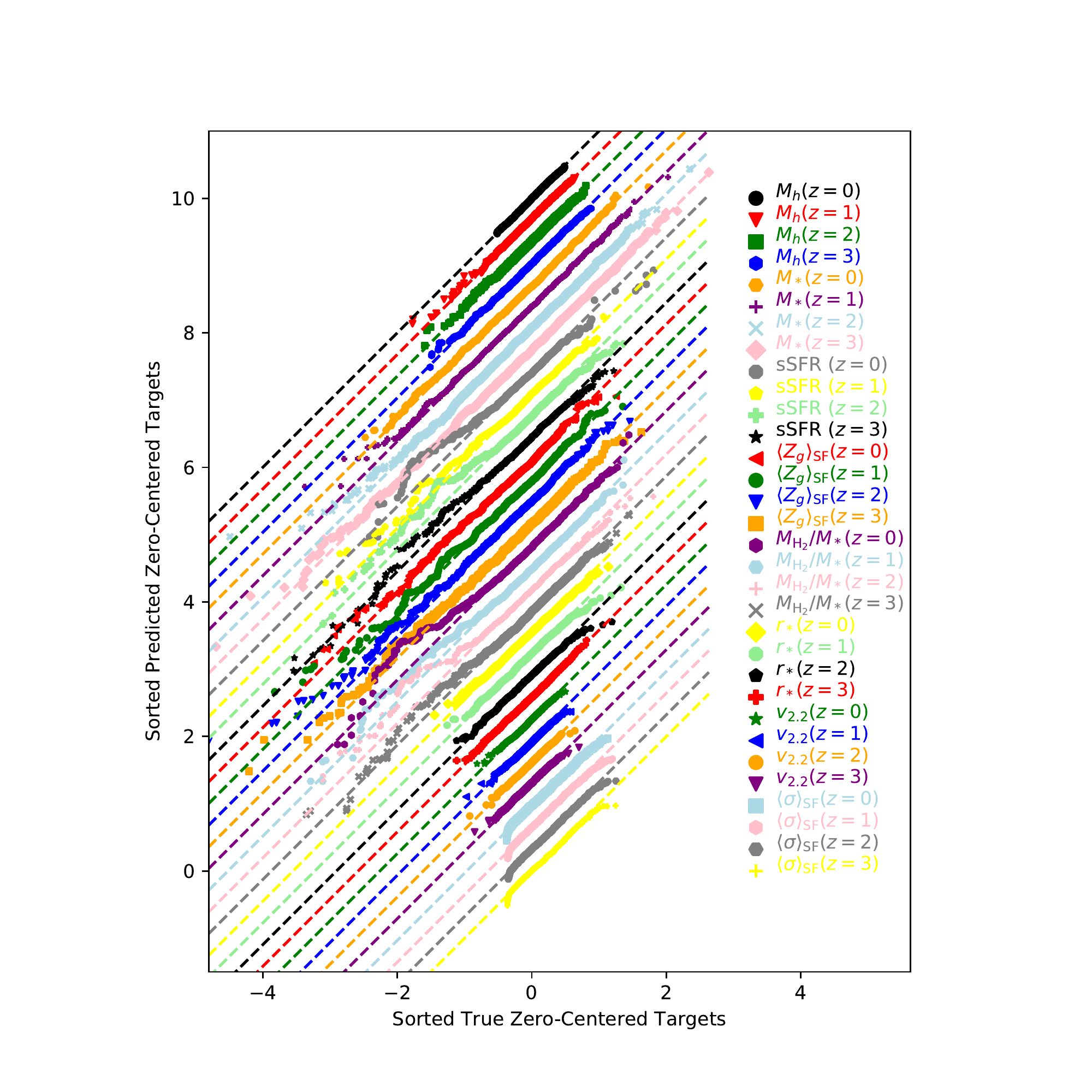}
\caption{QQ plot. As described in the text, for each target variable the values of the validation set are sorted and compared to the sorted values of the emulator's predictions (and both are centered to have zero median). A good match between the distribution of quantities predicted by the emulator and the true values corresponds to the points lying along the $y=x$ line. Each set of points and corresponding line is then offset vertically for visual clarity. The emulator predicts 8 different quantities, namely those shown in Equation \eqref{eq:targets} and Table \ref{tab:performance}, at 4 different redshifts. }
\label{fig:QQ}
\end{figure}

Aside from the residuals, the fit may be assessed with a quantile-quantile (QQ) plot. For a given target variable, the values of that variable in the validation set are sorted. The fit's predicted values for that variable, given the values of the input features in the validation set, are also sorted. The first value in each list, then the second value and so on, are paired with each other and plotted. If the predicted values and the actual values are drawn from the same distribution, this plot will appear as a line near $y=x$. To conserve space on the plot, these distributions are re-centered by subtracting the median of the true distribution from both the true and the predicted distributions. Additionally, since these quantites are all logarithmic when the emulator sees them, the values shown are also logarithmic. The values are not rescaled though, so the different dynamic ranges of each target are still visible in Figure \ref{fig:QQ}. For the most part, the samples do indeed fall close to the diagonal dashed lines, indicating good agreement between the predicted distributions and the actual distributions of the target variables. 

We can also use a few standard metrics to assess the fits. The first is the $R^2$ coefficient, defined as 
\begin{equation}
R^2 = 1 - \left(\sum_i (\psi_{\mathrm{val}, i} - \psi_{\mathrm{pred},i})^2  \right) / \left( \sum_i (\psi_{\mathrm{val},i} - \bar{\psi}_\mathrm{val})^2  \right)
\end{equation}
Here the sums extend over every element in the validation set, $\psi_{\mathrm{val},i}$ is the $i$th true value of the target quantity, $\psi_{\mathrm{pred},i}$ is the corresponding $\psi$ value predicted by the regression, and $\bar{\psi}_\mathrm{val}$ is the arithmetic mean of the $\psi_{\mathrm{val},i}$. By definition, if the model simply predicted the mean value of the training set (and the training and test sets were sufficiently large and drawn from the same distribution), $R^2=0$. If the prediction were perfect so that $\psi_{\mathrm{val},i} = \psi_{\mathrm{pred},i}$, $R^2=1$. Note that $R^2$ can be negative if the predictions are worse than simply predicting the average.

Another standard metric is to simply measure the coefficient of correlation between the test set and the predictions, i.e.
\begin{equation}
\rho = \frac{ \left(\sum_i (\psi_{\mathrm{pred},i} - \bar{\psi}_\mathrm{pred})(\psi_{\mathrm{val},i} - \bar{\psi}_\mathrm{val})  \right)}{ \left(    \sum_i (\psi_{\mathrm{pred},i} - \bar{\psi}_\mathrm{pred})^2 \sum_i(\psi_{\mathrm{val},i} - \bar{\psi}_\mathrm{test})    \right)^{1/2}}
\end{equation}
This coefficient, unlike $R^2$, is guaranteed to be between $-1$ and $1$, the latter implying a perfect match between the regression prediction and the test set, and the former implying a perfect anti-correlation between the two.

We are also particularly interested in the outlier fraction, namely how often the emulator makes a large error, which we define as a residual $\Delta = \psi_{\mathrm{pred},i} - \psi_{\mathrm{val},i}$ with absolute value greater than some quantity, e.g. $|\Delta|>0.3$ or $|\Delta|>1$ since these can cause problems for the MCMC. All of these quantities, along with the mean absolute error, i.e. the average of $|\Delta|$, are shown for each target variable for the emulator used in the MCMC. The performance metrics are all written such that lower values are better.

\begin{table}
\caption{Performance of the emulator.}
\begin{tabular}{l|llllll}
Variable & $1-R^2$ & AAE$^1$ &  MAE$^2$ & $1-\rho$ & $f_{|\Delta|>0.3}$ & $f_{|\Delta|>1.0}$ \\
\hline
\hline
$\mathbf{M_h}$ & & & & & & \\
$ (z=0)$ &    0.0041 &  0.0142 &  0.0114  &  0.0018 & 0.0& 0.0 \\
$ (z=1)$ &    0.0124&   0.0232 &  0.0161  &  0.0060 &0.0011&0.0 \\
$ (z=2)$ &    0.0181 &  0.0309 &  0.0215 &   0.0086 &  0.0028&0.0 \\
$ (z=3)$ &    0.0194 &  0.0369 &  0.0281  &  0.0095 &0.0033&0.0 \\
\hline
$\mathbf{M_*}$ & & & & & & \\
$(z=0)$ &   0.0347 &  0.0833 &  0.0509  &  0.0175 & 0.0386&0.00056\\
$ (z=1)$ &   0.0359 &  0.0894 &  0.0557 &   0.0179 &0.0454&0.00056\\
$ (z=2)$ &   0.0450 &  0.1055 &  0.0603 &   0.0227 &0.0616& 0.00448\\
$(z=3)$ &   0.0174  & 0.0852&   0.0588&    0.0086& 0.0319& 0.00056\\
\hline
$\mathbf{sSFR}$ & & & & & & \\
 $(z=0)$ &   0.1238&   0.0909 &  0.0492 &   0.0626 & 0.0588& 0.00056\\
 $(z=1)$ &   0.1062 &  0.0913 &  0.0537 &   0.0541 &0.0521& 0.00168\\
 $(z=2)$ &   0.1528&   0.1019 &  0.0521 &   0.0793 & 0.0773& 0.00336\\
 $(z=3)$ &   0.0532 &  0.0666&   0.0376&    0.0263&   0.0274&0.00056\\
\hline
$\mathbf{\langle Z_g\rangle_\mathrm{SF}}$ & & & & & & \\
$ (z=0)$ &     0.0590 &  0.0835&   0.0553 &   0.0296 &0.0308& 0.00056\\
$ (z=1)$ &     0.0495 &  0.0844 &  0.0539  &  0.0246& 0.0330&  0.00056\\
$ (z=2)$ &    0.0428 &  0.0866 &  0.0615 &   0.0210& 0.0347&0.00112\\
$ (z=3)$ &    0.0417 &  0.0857 &  0.0572  &  0.0205 &0.0358&  0.00056\\
\hline
$\mathbf{M_{\mathrm{H}_2}/M_*}$ & & & & & & \\
$ (z=0)$ &  0.0848&   0.1060 &  0.0644&    0.0424& 0.0650&  0.00168\\
$ (z=1)$ & 0.0906&   0.1035&   0.0635 &   0.0455 & 0.0644& 0.00392\\
$ (z=2)$ & 0.1413&   0.1147 &  0.0623  &  0.0726 & 0.0813&  0.00560\\
$ (z=3)$ & 0.0584&   0.0752 &  0.0485&    0.0289 &0.0280&0.001121\\
\hline
$\mathbf{r_*}$ & & & & & & \\
$(z=0)$ &   0.0907 &  0.0697 &  0.0486  &  0.0463&  0.0179&0.0\\
$(z=1)$ &   0.0833&   0.0587 &  0.0393 &   0.0423 & 0.01458&0.0\\
$(z=2)$ &   0.0806 &  0.0511  & 0.0346  &  0.0405 &0.0067& 0.00056\\
$(z=3)$ &   0.0468&   0.0425 &  0.0298  &  0.0234 & 0.0056&0.0\\
\hline
$\mathbf{v_{2.2}}$ & & & & & & \\
$(z=0)$& 0.0406 &  0.0234 &  0.0160 &   0.0201 &0.0& 0.0\\
$(z=1)$ &0.0364&   0.0238 &  0.0163  &  0.0179& 0.0& 0.0\\
$(z=2)$ &0.0340 &  0.0234 &  0.0161 &   0.0166 & 0.0 &0.0\\
$(z=3)$ &0.03033&  0.0230 &  0.0163  &  0.0149 &0.0& 0.0\\
\hline
${\langle\mathbf{\sigma}\rangle_\mathrm{SF}}$ & & & & & & \\
$(z=0)$& 0.0365 &  0.0446 &  0.0289 &   0.0182 & 0.0067&0.0\\
$(z=1)$ &0.03264&  0.0382&   0.0240 &   0.0162& 0.0061&0.0\\
$(z=2)$ &0.0319 &  0.0370 &  0.0238  &  0.0158 & 0.0050&0.0\\
$(z=3)$ &0.0361 &  0.0402 &  0.0271  &  0.0178 &0.0039&0.0\\
\hline
\end{tabular} \par
\label{tab:performance}
\bigskip
$^1$ Average Absolute Error. \\
$^2$ Median Absolute Error.
\end{table}

\clearpage

\end{document}